\documentclass[11pt]{article}
\pdfoutput=1 
\usepackage{cite}
\usepackage{graphicx}
\usepackage{caption}
\usepackage{refstyle}
\usepackage{amsmath}
\usepackage{amsthm}
\usepackage{amsmath}
\usepackage{amsfonts}
\usepackage{amssymb}
\usepackage{marvosym}
\usepackage{bbm}
\usepackage{amsmath,array}
\usepackage{mathtools}
\usepackage[all]{xy}
\usepackage{mathtools}
\usepackage[utf8]{inputenc}
\usepackage[english]{babel}
\usepackage[T3,T1]{fontenc}
\usepackage{fancyhdr}
\DeclareSymbolFont{tipa}{T3}{cmr}{m}{n}
\DeclareMathAccent{\invbreve}{\mathalpha}{tipa}{16}
\usepackage[left=1.7cm,right=1.7cm,top=2cm,bottom=2cm]{geometry}
\usepackage{setspace}
\usepackage{geometry}
\newtheorem{theorem}{Theorem}

 \def\mathbi#1{\textbf{\em #1}}
 
\begin{document}

\title{\LARGE On the Optimality of Gaussian Code-books for Signaling over a Two-Users Weak Gaussian Interference Channel}

\author{Amir K. Khandani\thanks{The author gratefully acknowledges the many helpful comments and suggestions of M. Costa, A. Gohari, and C. Nair on earlier versions of this article.}\\
{E\&CE Dept., Univ of Waterloo, Waterloo, Canada, khandani@uwaterloo.ca }}
%{\bf Readers are encouraged to first watch a summary video presentation at: https://cst.uwaterloo.ca/IC} }}
\maketitle

 \noindent 
{\bf Abstract:} This article establishes that the capacity region of the two-user weak Gaussian interference channel can be achieved using single-letter Gaussian codebooks.
It is further shown that the upper concave envelope of the achievable rate region can be attained using at most two time-sharing phases. In one phase, both users transmit simultaneously, while in the second phase, when present, only one user is active.
Furthermore, the boundary of the capacity region can be traversed continuously through incremental reallocations of power between the two messages transmitted by each user.
Finally, it is proven that the Han–Kobayashi achievable rate region with single-letter Gaussian codebooks attains the optimal boundary of the capacity region.

   \section{Introduction} \label{Intro}

  	\begin{figure}[h]
  	\centering
  	\includegraphics[width=0.45\textwidth]{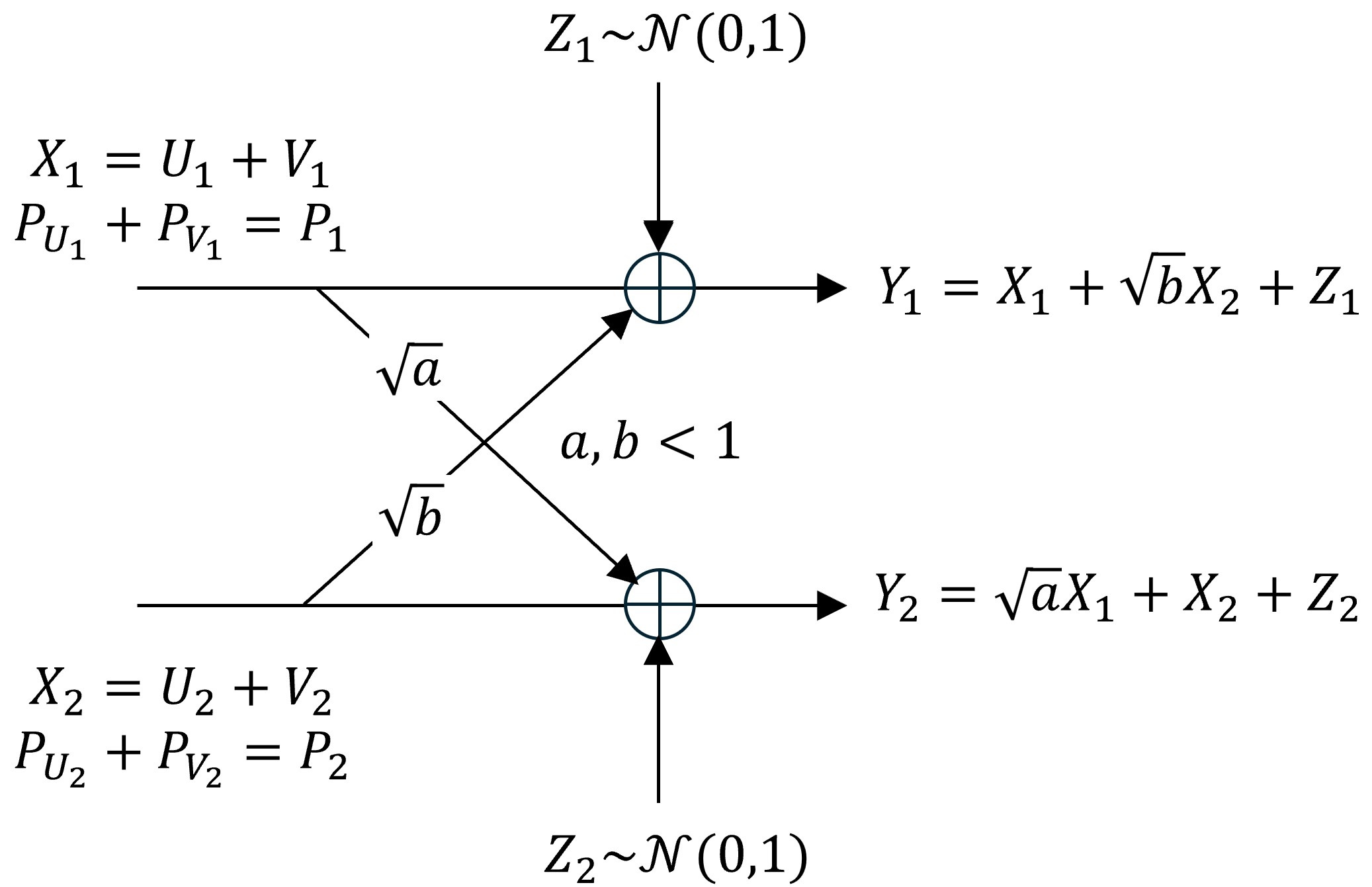}
  	\caption{Two users Gaussian Interference Channel (GIC) with  scalar inputs, $a<1$ and $b<1$.}
  	\label{Fig-G}
  \end{figure}
  Consider a two users Gaussian interference channel (see Fig.~\ref{Fig-G}) with inputs $X_1$, $X_2$ and outputs $Y_1$, $Y_2$, defined as
  \begin{eqnarray} \label{Eq1NN}
  	Y_1 & = & X_1+\sqrt{b}X_2+Z_1\\ \label{Eq1}
  	Y_2 & = & \sqrt{a} X_1+X_2+Z_2 
  \end{eqnarray}
  where $a,b<1$, $Z_1$, $Z_2$ are additive white Gaussian noise  of zero  mean and unit variance. \underline{Note}: All discussions in this article, except for Theorem~\ref{S4NT},  are formulated in terms of the message construction used in establishing Han-Kobayashi (HK) achievable rate region. Theorem \ref{S4NT} establishes the optimality of Gaussian code-books in the general case.  It will be shown that, subject to (possible) time-sharing between two phases,  the single-letter HK rate region is optimum. 
  
 Relying on the message construction used in establishing Han-Kobayashi (HK) achievable rate region, we have 
  \begin{eqnarray} \label{Eq1NNN}
  	X_1 & = & U_1+V_1\\ \label{Eq1N}
  	X_2 & = & U_2+V_2
  \end{eqnarray}
where messages $U_1$ and $U_2$ are public, while $V_1$ and $V_2$ are private\footnote{Although the formulation considers only one public and one private message per user, the arguments extend to a more general setting in which each user is allowed multiple public and/or multiple private messages (see Remark 1}. Theorem~\ref{S4NT} shows arguments establishing Gaussianity of code-books apply to the general formulation (relying on vector inputs without restricting the messages to public and private pairs). 

 The maximization of the weighted sum-rate can be expressed as
  \begin{eqnarray} 
\label{13-0}
  	\mbox{Maximize:} \qquad
  	R_1+\mu R_2
  	&=&
  	R_{U_1}+R_{V_1}
  	+
  	\mu\left(R_{U_2}+R_{V_2}\right)
  	\\ \label{13-1}
  	\mbox{Subject to:} \qquad
  	P_{U_1}+P_{V_1}
  	&=&
  	P_1
  	\\   \label{13-2}
  	P_{U_2}+P_{V_2}
  	&=&
  	P_2.
  \end{eqnarray}
  Regardless of the codebook structure (scalar or vector), the random-coding distributions employed, or the decoding strategy used to recover $U_1$ and $U_2$ (i.e., joint or successive decoding), we have
  \begin{eqnarray} \label{NNEq6ttV}
  	R_{V_1}  & \leq  & I(V_1;Y_1|U_1,U_2) \\  \label{NNEq7ttV}
  	R_{V_2}  & \leq & I(V_2;Y_2|U_1,U_2) \\  \label{NNEq8tt}
  	R_{U_1}  & \leq  & I(U_1;Y_1|U_2) \\  \label{NNEq7tt}
  	R_{U_1}  & \leq  & I(U_1;Y_2|U_2) \\  \label{NNEq8ttt}
  	R_{U_2}  & \leq  & I(U_2;Y_1|U_1) \\  \label{NNEq9tt}
  	R_{U_2}  & \leq  & I(U_2;Y_2|U_1) \\ \label{NNEq6}
  	R_{U_1}+R_{U_2}  & \leq  & I(U_1,U_2;Y_1) \\  \label{NNEq7}
  	R_{U_1}+R_{U_2} & \leq  & I(U_1,U_2;Y_2).
  	\label{V4Eq1}
  \end{eqnarray}
  
  For given values of 	$P_{U_1}$, $P_{V_1}$, $P_{U_2}$, $P_{V_2}$ 
  satisfying \ref{13-1} and  \ref{13-2}, maximizing $	R_{U_1}+R_{V_1}
  +
  \mu\left(R_{U_2}+R_{V_2}\right)$ in \ref{13-0} subject to \ref{NNEq6ttV} to \ref{NNEq7} is a linear program for which the number of constraints among \ref{NNEq6ttV} to \ref{NNEq7} satisfied with equality (active constraints)  is at least equal to the number of non-zero rate values among $R_{U_1}$, $R_{V_1}$, $R_{U_2}$, $R_{V_2}$ (basic variables). Non-zero rate values among $R_{U_1}$, $R_{V_1}$, $R_{U_2}$, $R_{V_2}$ can be computed in terms of the mutual information quantities by solving the system of linear equations corresponding to the active constraints. 

Solving the optimization problem in \ref{13-0} to \ref{NNEq6ttV} entails:
(a) For each user, partitioning the available power between the public and private messages, referred to as power allocation {\em over messages}.
(b) Determining the optimal density function for each message codebook.
(c) Identifying the active constraints among \ref{NNEq6ttV} to \ref{NNEq7}.
(d) Determining the encoding and decoding procedures for each user.
The term {\em coding strategy}, or simply {\em strategy}, refers to the selection of items (a)–(d) for each user at a given point on the boundary.
The optimization problem in \ref{13-0}  to \ref{NNEq7}  pertains to a single time-sharing phase. Forming the upper concave envelope requires time sharing among multiple phases, together with optimizing power allocation {\em over time}, which governs how each user’s power is distributed across the time-sharing phases to maximize the overall weighted sum-rate. Note that each time-sharing phase is equipped with its own coding strategy, designed to maximize the contribution of that phase to the overall weighted sum-rate. This work adopts a divide-and-conquer approach consisting of: (1) decomposing the above complex problem into simpler sub-problems, (2) solving these sub-problems sequentially, and (3)  combining the results to draw conclusions about the underlying random coding strategies. 
  
\begin{figure}[h]
	\centering
	\includegraphics[width=0.5\textwidth]{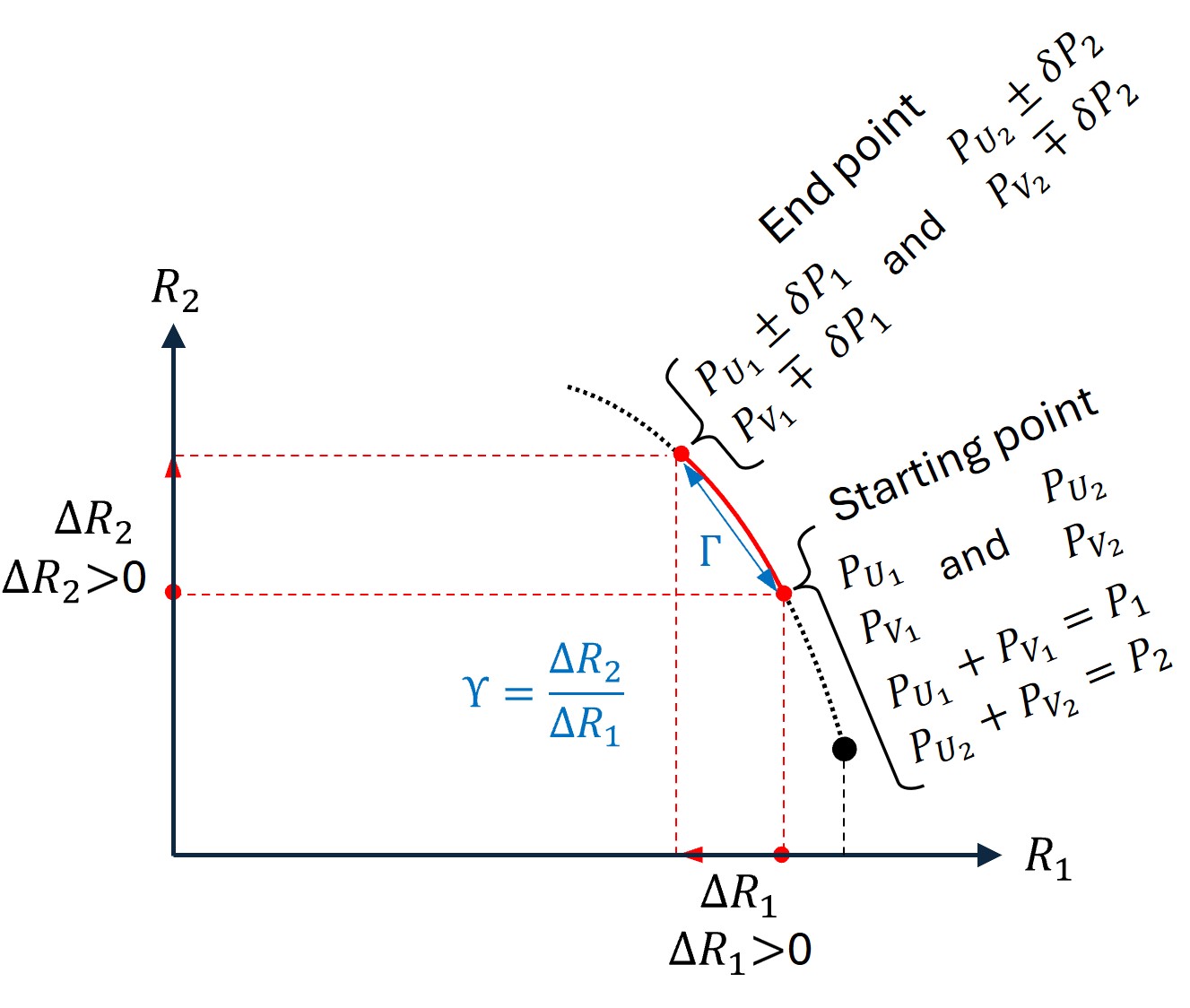}
	\caption{An example for power reallocation and its corresponding step along the boundary. }
	\label{Inc}
\end{figure}

In addition, it is of interest to characterize the points on the boundary of the capacity region. Owing to the large number of optimization parameters, this is a challenging problem.
To simplify the analysis, we focus on traversing the boundary by varying the power allocation between the public and private messages, referred to hereafter as {\em power reallocation} (over messages). Each step begins at a point on the boundary. The power allocation over messages is then perturbed, and the optimal codebooks corresponding to the new allocation are determined such that the step terminates at another point on the boundary.
The power reallocation values corresponding to such a step satisfy
\begin{eqnarray} \label{Eq0PZ}
	(P_{U_1},P_{V_1}) & \xRightarrow[\text{}]{\text{Power reallocation:\,} (\kappa_1,\eta_1)} & (P_{U_1}+\kappa_1,P_{V_1}+\eta_1):~ \kappa_1+\eta_1=0 \\ \label{Eq1PZ}
(P_{U_2},P_{V_2}) &\xRightarrow[\text{}]{\text{Power reallocation:\,} (\kappa_2,\eta_2)} & (P_{U_2}+\kappa_2,P_{V_2}+\eta_2):~\kappa_2+\eta_2=0. 
\end{eqnarray}
With some misuse of notations, hereafter power reallocation vectors are denoted  as   
\begin{eqnarray} \label{Eq0PZ2}
(\delta P_{1},\delta P_{2})~\mbox{where}~~  
\delta P_{1}=|\kappa_1|=|\eta_1|,~~\delta P_{2}=|\kappa_2|=|\eta_2|.
\end{eqnarray}
In other words, $\delta P_{1}$ denotes the increase 
in the power of $U_1$ or $V_1$, depending on which of the two has a higher power at the end point vs. the starting point, and likewise for $\delta P_{2}$ in relation to $U_2$ and $V_2$.
Figure~\ref{Inc} depicts an example where notations $\pm$ vs. $\mp$ are used to emphasize that the signs of $\delta P_{1}$ and $\delta P_{2}$ depend on the step and power reallocation is zero-sum. Power reallocation vector is selected to guarantee the solution achieving each end point is unique. To achieve the latter criterion, while moving continuously along the boundary,  power reallocation vector is selected relying on a notation of admissibility called Pareto minimal. Referring to Fig.~\ref{Inc}, for a given a power reallocation vector $(\delta P_1,\delta P_2)$, the following measures are used in arguments related to the next point on the boundary (end point):  Slope of the step $\mathbf{\Upsilon}$ and length of the step $\mathbf{\Gamma}$ (see Fig.~\ref{Inc}). 
%Given $(\delta P_1,\delta P_2)$, random coding for the end point  maximizes the length  of the step, i.e., $\mathbf{\Gamma}$, over all possible values of the slope $\mathbf{\Upsilon}$ as depicted in Fig.~\ref{Inc}.  

Hereafter,   $U_1$, $U_2$,  $V_1$, $V_2$ are called core random variables. Linear combinations of core random variables appearing in  mutual information terms are called compound random variables. Unlike core random variables which are independent, compound random variables are generally dependent on each other. We will see that, although dependent,  compound random variables can vary individually. This means the conditional density of any compound random variable conditioned on the rest has a non-zero entropy. 

\section{Literature Survey}

The problem of Gaussian interference channel has been the subject of numerous outstanding prior works, paving the way to the current point and moving beyond. A subset of these works, reported in \cite{GIC1} to \cite{GIC38}, are briefly discussed in this section.  A more complete and detailed literature survey will be provided in subsequent revisions of this article

Reference~\cite{GIC1} discusses degraded Gaussian interference channel (degraded means one of the two receivers is a degraded version of the other one) and presents multiple bounds and achievable rate regions. Reference~\cite{GIC2} studies the capacity of two users GIC for the class of strong interference and shows the capacity region is at the intersection of two MAC regions, consistent with the current article.  Reference~\cite{GIC3}  establishes optimality for two extreme points in the achievable region of the general two users GIC. \cite{GIC3} also proves that the class of degraded Gaussian interference channels is equivalent to the class of Z (one-sided) interference channels. 

References~\cite{GIC4} to \cite{GIC6} present achievable rate regions for interference channel. In particular, \cite{GIC4} presents the well-known Han-Kobayashi (HK) achievable rate region.  HK rate region coincides with all results derived previously (for Gaussian two users GIC), and is shown to be optimum for the class of weak two users GIC in the current article. References \cite{GIC7}\cite{New1} have further studied the HK rate region.  Reference \cite{New1} shows that HK achievable rate region is strictly sub-optimum for a class of discrete interference channels. 

References~\cite{GIC9} to \cite{GIC15} have studied the problem of outer bounds for the interference channel.  Among these, \cite{GIC11}\cite{GIC12}\cite{GIC13} have also provided optimality results in some special cases of weak two users GIC.  

References~\cite{GIC16}\cite{GIC17} have studied the problem of interference channel with common information. References~\cite{GIC18} to \cite{GIC20} have studied the problem of interference channel with cooperation between transmitters and/or between receivers. 
References~\cite{GIC21}\cite{GIC22} have studied the problem of interference channel with side information. Reference~\cite{GIC23} has studied the problem of interference channel assuming cognition, and  reference~\cite{GIC24} has studied the  problem assuming cognition, with or without secret messages. 

Reference \cite{GIC25} has found the capacity regions of vector Gaussian interference channels for classes of very strong and aligned strong interference. \cite{GIC25} has also generalized some known results for 
sum-rate of scalar Z interference, noisy interference, and mixed interference to the case of vector channels. Reference \cite{GIC26} has addressed the sum-rate  of the parallel Gaussian
interference channel. Sufficient conditions are derived
in terms of problem parameters (power budgets and channel coefficients) such
that the sum-rate can be realized by independent transmission across sub-channels while treating interference as noise, and corresponding optimum power allocations are computed. 
Reference \cite{GIC27} studies a Gaussian interference network where each message is encoded by a single transmitter and is aimed at a single receiver. Subject to feeding back the output from receivers to their corresponding transmitter, efficient strategies
are developed based on the discrete Fourier transform signaling. 

Reference \cite{GIC28} computes the capacity of interference channel within one bit. 
References \cite{GIC29}\cite{GIC30} study the impact of interference in GIC.  \cite{GIC30} shows that treating interference as noise in two users GIC achieves the closure of the capacity
region to within a constant gap, or within a gap that scales as O(log(log(.)) with signal to noise ratio. Reference \cite{GIC31} relies on game theory  to define
the notion of a Nash equilibrium region of the interference
channel, and characterizes the Nash equilibrium region for: (i) two users
linear deterministic interference channel in exact form, and (ii) two users GIC within 1 bit/s/Hz in an approximate form. 

Reference \cite{GIC32} studies the problem of two users GIC based on a sliding window superposition coding scheme. 

References \cite{GIC33} and  \cite{GIC34}, independently, introduce the new concept of non-unique decoding as an intermediate alternative to ``treating interference as noise'', or ``canceling interference''.  Reference~\cite{GIC35} further studies the concept on non-unique decoding and proves that (in all reported cases) it can be replaced by a special joint unique decoding without penalty.

Reference~\cite{GIC36} studies the degrees of freedom of the K-user
Gaussian interference channel, and, subject to a mild sufficient condition on the channel
gains, presents an expression  for the degrees of freedom of the scalar interference
channel as a function of the channel matrix.

Reference~\cite{GIC37} studies the problem of state-dependent Gaussian interference channel, where two receivers are affected by scaled versions of the same state. The state
sequence is  (non-causally) known at both transmitters, but not at receivers. 
Capacity results are established (under certain conditions on channel parameters) in the very strong, strong, and weak interference regimes. For the weak regime, the sum-rate is computed. Reference~\cite{GIC38} studies the problem of state-dependent Gaussian interference channel under the assumption of 
correlated states, and characterizes  (either fully or partially) the capacity region  or the sum-rate under various channel parameters. 

Reference \cite{Amin} settles the noiseberg conjecture\cite{Amin2} regarding the Han-Kobayashi region of the Gaussian Z-Interference channel
with Gaussian signaling.

\section{Article Organization}
\label{MRES}
Section \ref{S4N} establishes that Gaussian code-books are optimum for general vector messages. The rest of the article focuses on familiar HK formulation, i.e., public/private messages for each user. 
Section \ref{S4} is devoted to determining the optimal densities for the compound random variables. Section \ref{S41} shows that at least one of the receivers can rely on successive decoding. This allows formulating the constraints on rate values in terms of mutual information terms corresponding to additive noise channels. Relying on this formulation, 
in Section \ref{S4.1}, by invoking the Linear Independence Constraint Qualification (LICQ), it is shown that Gaussian vector densities for the compound random variables maximize the weighted sum-rate. This result is then used in Section \ref{S4.2} to establish a degradedness property for the weak two-user Gaussian interference channel (GIC).  The degradedness property is used to determine the set of active constraints among 
\ref{NNEq6ttV} to \ref{V4Eq1} and the corresponding decoding orders. 
Relying on the resulting formulation, Section \ref{S4.3} shows that single-letter Gaussian densities are optimal for the compound random variables.

Section \ref{S5} is devoted to determining the optimal densities for the core random variables.
In Section \ref{S5.1}, it is shown that the core and compound random variables are related through a full-rank system of linear equations. Consequently, by invoking the result of Section \ref{S4.3} concerning the optimality of single-letter Gaussian densities for the compound random variables, it is concluded that single-letter Gaussian codebooks are also optimal for the core random variables.
One of the principal challenges in characterizing the capacity region of the two-user weak Gaussian interference channel is determining how to allocate the total power available to each user among the constituent channels associated with the time-sharing phases, and if embedding information in the selection of  time-sharing phases is required. This problem is commonly referred to as determining the {\em upper concave envelope} of the achievable rate region.
Section \ref{S5.2} shows that the number of constituent channels (time-sharing phases) required to achieve the upper concave envelope is at most two. In one phase, both users are active, while in a second phase, if present, only a single user is active. This result is consistent with \cite{Amin}, where the Han–Kobayashi achievable rate region \cite{GIC4} with Gaussian inputs was optimized for the Z-interference channel. 
Theorem \ref{Th8N} in Section \ref{S5.2} shows that embedding information in the selection of  time-sharing phase does not improve the weighted sum-rate.  
Section \ref{S5.3} establishes that the single-letter Gaussian densities used in the construction of the codebooks are zero-mean. Consequently, without loss of optimality, the optimization problem can be restricted to zero-mean densities.

Section \ref{ABC1} shows that the boundary of the capacity region can be traversed continuously through incremental reallocations of power between the public and private messages.

Converse results are established in Section~\ref{sec4p}.

Finally, Section~\ref{sec4} shows that the solution of the Han–Kobayashi achievable rate region with single-letter Gaussian random codebooks attains the optimal boundary of the capacity region.

   \section{Optimality of Gasussain Code-books for General Vector Inputs} \label{S4N}
\begin{figure}[h]
   	\centering
   	\includegraphics[width=0.45\textwidth]{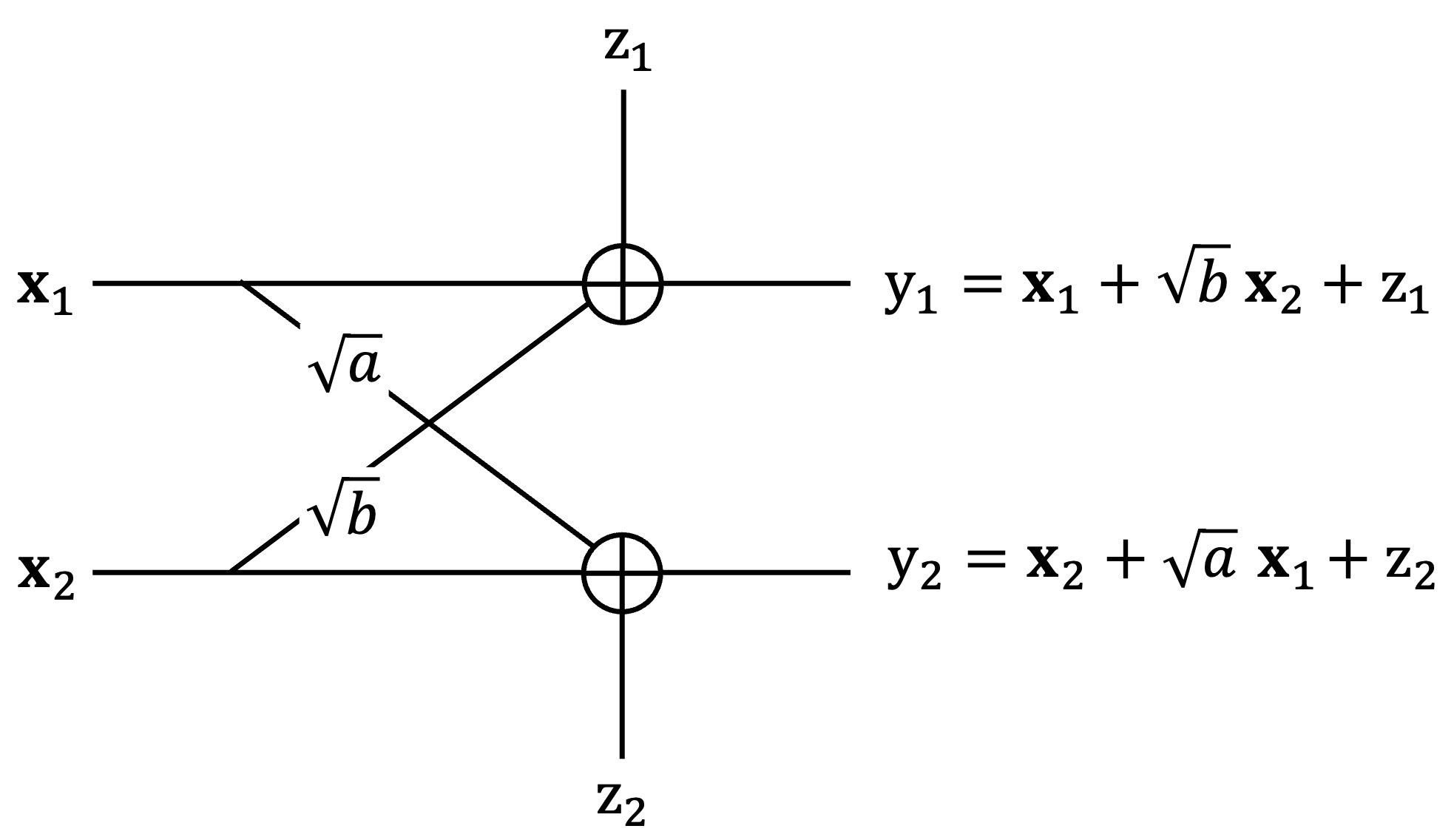}
   	\caption{Two users Gaussian Interference Channel (GIC) in  general (vector) form,  $\mathtt x_1$, $\mathtt x_2$ are independent of each other and $\|\mathtt x_1\|^2$,  $\|\mathtt x_2\|^2$ are fixed.}
   	\label{S4NF}
   \end{figure}

      Consider the channel in Fig.~\ref{S4NF} where  $\mathtt x_1$, $\mathtt x_2$ are vector inputs of power values $\|\mathtt x_1\|^2$,  $\|\mathtt x_2\|^2$, respectively,  $\mathtt z_1$ and $\mathtt z_2$ are additive Gaussian  noise vectors.  
   
   In continuous optimization, an important question is determining when first-order optimality conditions are necessary. For constrained problems, this issue is addressed through Constraint Qualifications (CQs), which ensure that the Karush--Kuhn--Tucker (KKT) conditions hold at any local optimum. One of the most fundamental CQs is the Linear Independence Constraint Qualification (LICQ). LICQ requires that the gradients of the active inequality constraints, together with the gradients of the equality constraints, be linearly independent~\cite{LICQ1,LICQ2,LICQ3,LICQ4,LICQ5}.  Relying on LICQ condition, Theorem~\ref{S4NT} establishes that, Gaussian densities for $\mathtt x_1$, $\mathtt x_2$ maximize the weight sum-rate $\mathbi{R}_1 +\mu 	\mathbi{R}_2$ where 
      \begin{eqnarray} \label{Eq201N}
   	   	\mathbi{R}_1  & = & H_{\mathtt{y}_1}(\mathbf x_1+\sqrt{b} \mathbf x_2+\mathbf z_1)-H_{\mathtt{n}_1}( \sqrt{b} \mathbf x_2+\mathbf z_1) \\ \label{Eq202N}
   	   	\mathbi{R}_2  & = & H_{\mathtt{y}_2}( \mathbf x_2+\sqrt{a}\mathbf x_1+\mathbf z_2)-H_{\mathtt{n}_2}( \sqrt{a} \mathbf x_1+\mathbf z_2).
   	   \end{eqnarray}

         \begin{theorem} 	\label{S4NT} 
Fixing $\|\mathtt x_1\|^2$, $\|\mathtt x_2\|^2$, any (local/global) optimum solution  maximizing $\mathbi{R}_1 +\mu 	\mathbi{R}_2$ should rely on Gaussian densities for  $\mathtt x_1$, $\mathtt x_2$. 
   \end{theorem}	
\begin{proof} 
See Appendices \ref{App1} and \ref{LIC2}. 
\end{proof} 
   \section{Optimality of Gasussain for Compound Public/Private (HK) Messages} \label{S4}
   \subsection{Successive Decoding of Public Messages}  \label{S41}
In this work, the sequence of arguments used throughout several proofs relies on the following key observation. A maximizing solution to the optimization problem defined by \ref{13-0} to \ref{NNEq7} cannot satisfy all the following conditions with strict inequality
\begin{eqnarray}  \label{E-19}
 R_{U_1} & < & I(U_1;Y_1) \\ \label{E-20}
  R_{U_1} & < & I(U_1;Y_2) \\  \label{E-21}
   R_{U_2} & < & I(U_2;Y_1) \\ \label{E-22}
  R_{U_2} & < & I(U_2;Y_2).
\end{eqnarray}
The reason is that, if \ref{E-19} to \ref{E-22} are all strict inequalities, then the messages $U_1$ and $U_2$ can each be decoded at both receivers, $Y_1$ and $Y_2$, while treating remaining messages as  noise. Since the inequalities in \ref{E-19} to \ref{E-22} are strict, at least one of the rates, $R_{U_1}$ or $R_{U_2}$, can be increased without violating the  decodability constraint (as the fist layer in successive decoding at the respective receiver). Therefore, such a point cannot be a maximizing solution. Consequently, at least one of the inequalities in \ref{E-19} or \ref{E-22} must hold with equality.  In particular, if
$R_{U_2}=I(U_2;Y_1)$, then $U_2$ forms the first layer in a successive decoding procedure at the receiver $Y_1$. In this case,  the inequality sign in \ref{E-21} is replaced with equality --- this does not entail \ref{E-19}, \ref{E-20}, \ref{E-22} will be satisfied at the same time.  
Under this condition, as long as decodablity of $(U_1,U_2)$ at $Y_1$ is concerned, $R_{U_1}$ can incrases such that the inequality $R_{U_1}\leq  I(U_1;Y_1|U_2)$ is changed to
\begin{equation} 
	R_{U_1}=I(U_1;Y_2|U_2). 
	\label{E22}
\end{equation} 
Theorem \ref{V4Th5z} shows that indeed the conditions $R_{U_2}=I(U_2;Y_1)$ and $R_{U_1}=I(U_1;Y_2|U_2)$ can be both satisfied without violating the condition for decodability of $(U_1,U_2)$ at $Y_2$ --- by using joint decoding of $(U_1,U_2)$ at $Y_2$.
Equation \ref{E22} implies that, after $U_2$ has been decoded and removed, $U_1$ can be recovered at $Y_1$ through successive decoding. The same reasoning subsequently applies to the decoding of $V_1$ at $Y_1$.
The following theorem formalizes the above arguments. It further establishes that, owing to the independence of the messages $U_1$, $V_1$, $U_2$, $V_2$, the rate associated with each layer in a successive decoding procedure can be expressed as the mutual information of an additive-noise channel whose noise is independent of the channel input. 

\emph{Above arguments  are subsequently used in Section~\ref{S4.1} to establish the first result concerning Gaussianity of the codebooks. This key result acts as the cornerstone  for what follows.
  \underline{Note that} these  arguments (as well as Theorem \ref{V4Th5z}) rely solely on the independence of the messages, and apply equally to vector messages and vector codebooks.}
      \begin{theorem} 	\label{V4Th5z} 
   	In at least one of the receivers,  public messages  can be recovered using successive decoding. 
   \end{theorem}	

\begin{proof}   
	For simplicity, without loss of generality, proof is formulated in terms of scalar code-books.  Generalization to vector code-books follows noting that the only property used in establishing the results is independence of messages $U_1$, $V_1$, $U_2$, $V_2$,  which is also valid for vector messages  $\mathtt{u}_1$, $\mathtt{v}_1$, $\mathtt{u}_2$ and $\mathtt{v}_2$.
	
	Consider the linear programming problem of maximizing
	$R_{U_1}+R_{V_1}+\mu R_{U_2}+\mu R_{V_2}$
	subject to the constraints in \ref{NNEq6ttV} to \ref{NNEq7}. Since the objective function involves four variables, at least four of the constraints must be active at the optimum solution, i.e., satisfied with equality.
	The mutual information terms appearing on the right-hand sides of the active constraints represent rates achievable through successive decoding over additive-noise channels. Constraints \ref{NNEq6ttV} and \ref{NNEq7ttV} should be active, otherwise $R_{V_1}$  and/or  $R_{V_2}$ could be increased without affecting constraints in \ref{NNEq8tt} to \ref{NNEq7}. This means at least two of the constraints in \ref{NNEq8tt} to \ref{NNEq7} should be active. If any of the constraints 
	\ref{NNEq8tt} to \ref{NNEq9tt}	is active,  it entails  successive decoding at a respective receiver, and proof is complected. Otherwise, \ref{NNEq6} and \ref{NNEq7} should be both active, i.e., 
	\begin{eqnarray} \label{RE1p}
		R_{U_1}+R_{U_2}  & = &  I(U_1,U_2;Y_1)\,=\,I(U_2;Y_1)+I(U_1;Y_1|U_2)  \\ \label{RE2p}
		& = &   I(U_1,U_2;Y_2)\,=\,I(U_2;Y_2)+I(U_1;Y_2|U_2) \\ \label{RE3p}
		& = &  I(U_1,U_2;Y_1)\,=\,I(U_1;Y_1)+I(U_2;Y_1|U_1) \\ \label{RE4p}
		& = &  I(U_1,U_2;Y_2)\,=\,I(U_1;Y_2)+I(U_2;Y_2|U_1).
	\end{eqnarray} 
	We need to show that, in addition to \ref{NNEq6} and \ref{NNEq7}, at least one of the constraints \ref{NNEq8tt} to \ref{NNEq9tt} becomes active. 
	Let us assume \ref{NNEq8tt} to \ref{NNEq9tt} are satisfied with strict inequality.  Combining \ref{NNEq8tt}, \ref{NNEq7tt}, \ref{NNEq8ttt}, \ref{NNEq9tt} with \ref{RE1p}, \ref{RE2p}, \ref{RE3p}, \ref{RE4p}, respectively, would result in
	%   	entail at least one of the constraints in \ref{NNEq8tt} to \ref{NNEq9tt} should be satisfied with equality, indicating successive decoding at one of receivers. 
	\begin{eqnarray} \label{R27p}
		R_{U_2}~>~I(U_2;Y_1)~~\mbox{and}~~\ref{RE1p} & \rightarrow & R_{U_1}=I(U_1;Y_1|U_2)-\varkappa_1,~ R_{U_2}=I(U_2;Y_1)+\varkappa_1~\mbox{where}~\varkappa_1>0\\ \label{R28p}
		R_{U_2}~>~I(U_2;Y_2)~~\mbox{and}~~\ref{RE2p} & \rightarrow &
		R_{U_1}=I(U_1;Y_2|U_2)-\varkappa_2,~ R_{U_2}=I(U_2;Y_2)+\varkappa_2 ~\mbox{where}~\varkappa_2>0\\ \label{R29p}
		R_{U_1}~>~I(U_1;Y_1)~~\mbox{and}~~\ref{RE3p} & \rightarrow & 
		R_{U_2}=I(U_2;Y_1|U_1)-\varkappa_3,~ R_{U_1}=I(U_1;Y_1)+\varkappa_3 ~\mbox{where}~\varkappa_3>0\\ \label{R30p}
		R_{U_1}~>~I(U_1;Y_2)~~\mbox{and}~~\ref{RE4p} & \rightarrow & 
		R_{U_2}=I(U_2;Y_2|U_1)-\varkappa_4,~ R_{U_1}=I(U_1;Y_2)+\varkappa_4~\mbox{where}~\varkappa_4>0.
	\end{eqnarray}
	Next, it will be shown that at least one of the inequalities in \ref{R27p}, \ref{R28p}, \ref{R29p}, \ref{R30p}  will be satisfied with equality. 
	The value of $R_{U_1}+\mu R_{U_2}$ corresponding to  \ref{R27p}, \ref{R28p}, \ref{R29p}, \ref{R30p} is, respectively, equal to
	\begin{eqnarray} \label{R31}
			R_{U_2}~>~I(U_2;Y_1)~~& \rightarrow &~~
		R_{U_1}+\mu R_{U_2}  =  I(U_1;Y_1|U_2)+\mu\, I(U_2;Y_1)-\varkappa_1\,(1-\mu)  \\ \label{R32}
		R_{U_2}~>~I(U_2;Y_1)~~& \rightarrow &~~
		R_{U_1}+\mu R_{U_2} =  I(U_1;Y_2|U_2)+\mu\, I(U_2;Y_2)-\varkappa_2\,(1-\mu) \\ \label{R33}
				R_{U_1}~>~I(U_1;Y_1)~~& \rightarrow &~~
		R_{U_1}+\mu R_{U_2}  =  I(U_1;Y_1)+\mu\, I(U_2;Y_1|U_1)-\varkappa_3\,(\mu-1) \\ \label{R34}
						R_{U_1}~>~I(U_1;Y_2)~~& \rightarrow &~~
		R_{U_1}+\mu R_{U_2}  =  I(U_1;Y_2)+\mu\, I(U_2;Y_2|U_1)-\varkappa_4\,(\mu-1).
	\end{eqnarray}
 For $\mu<1$, it follows that   either $\varkappa_1=0$,  $\varkappa_2\geq 0$, or
 $\varkappa_1\geq 0$,  $\varkappa_2=0$,  would result in a higher  value for $R_{U_1}+\mu R_{U_2}$. Let us focus on $\varkappa_1=0$ which is achieved  for 
 $R_{U_1}=I(U_1;Y_1|U_2)$ and $R_{U_2}=I(U_2;Y_1)$, corresponding to successive deocding of $U_2$ followed by $U_1$ at $Y_1$. Note that for  weighted sum-rate $R_{U_1}+\mu R_{U_2}$ is maximized by optimizing power allocation satisfying $\varkappa_1=0$. $\varkappa_1=0$ is Likewise, for $\varkappa_2=0$, we have 
 $R_{U_1}=I(U_1;Y_2|U_1)$ and $R_{U_2}=I(U_2;Y_2)$, corresponding successive decoding of $U_2$ followed by $U_1$ at $Y_2$.       \end{proof}

 \vspace{0.2cm}
 \noindent 
 {\bf Remark 1:}  In the case of allowing for multiple public and/or private messages, relying on successive decoding,  entropy terms in expressions in \ref{E36} to \ref{E41} change to conditional entropy and arguments remain valid. $\blacksquare$
 
 \vspace{0.2cm}
 \noindent 
 {\bf Remark 12} Let us consider the case of  $\varkappa_1=0$ in Theorem~\ref{V4Th5z}.  Corresponding  constraints on rates are
 \begin{eqnarray} \label{E36}
 	R_{U_2}& = & I(U_2;Y_1)\,=\,H(U_1+V_1+\sqrt{b} U_2+\sqrt{b} V_2+Z_1)-H(U_1+V_1+\sqrt{b} V_2+Z_1) \\   \label{E37}
 	R_{U_1} & = & I(U_1;Y_1|U_2)\,=\,H(U_1+V_1+\sqrt{b} V_2+Z_1)-H(V_1+\sqrt{b} V_2+Z_1) \\  \label{E38}
	R_{V_2}& = & I(V_2;Y_2|U_1,U_2)\,=\, 
	H(V_2+\sqrt{a} U_1+\sqrt{a} V_1+Z_2)-H(\sqrt{a} U_1+\sqrt{a}V_1+Z_2)  \\  \label{E39}
 	R_{V_1}& = & I(V_1;Y_1|U_1,U_2)\,=\, H(V_1+\sqrt{b} V_2+Z_1)-H(\sqrt{b} V_2+Z_1) \\  \label{E40}
 	 R_{U_1}+R_{U_2}& = &  I(U_1,U_2;Y_1)=\,H( U_1+\sqrt{b} U_2+V_1+\sqrt{b}V_2+Z_1)-H(V_1+\sqrt{b}V_2+Z_1) \\ \label{E41}
 R_{U_1}+R_{U_2}& \leq &  I(U_1,U_2;Y_2)=\,H(\sqrt{a} U_1+U_2+\sqrt{a}V_1+V_2+Z_2)-H(\sqrt{a}V_1+V_2+Z_2).
 \end{eqnarray}
 Expressions  \ref{E36} to \ref{E40} correspond to active constraints which will be involved in verifying LICQ.  Noting the dual linear program, having the  slack variable in \ref{E41} equal to zero, if possible, increases  the objective function.  This means the optimum bandwidth/power allocation aims to satisfy \ref{E41} with equality. As a result, \ref{E41} will be also considered in verifying LICQ. $\blacksquare$

   \subsection{Optimality of Gaussian for Multi Letter (Vector) Compound Variables}  \label{S4.1}

Theorem \ref{V4Th5p} establishes  that the LICQ condition is satisfied  for  maximization problem governing the weighted sum-rate. 
   \begin{theorem}
A necessary condition for optimality of a set of vector code-books is that the first-order variation of the weighted sum-rate vanishes; and this condition can be satisfied  only if  code-books are Gaussian.
% \begin{eqnarray} 	\label{ApEq1ppn}
% 	\mathfrak{r}_{\mathtt{u}_2}
% 	& = &
% 	H_1( \mathtt{u}_1+\mathtt{v}_1
% 	+\sqrt{b}\mathtt{u}_2+\sqrt{b}\mathtt{v}_2+\mathtt{z}_1)
% 	-
% 	H_3( \mathtt{u}_1
% 	+\mathtt{v}_1+\sqrt{b}\mathtt{v}_2+\mathtt{z}_1)
% 	\\ \label{ApEq2ppn}
% 	\mathfrak{r}_{\mathtt{u}_1}
% 	& = &
% 	H_3( \mathtt{u}_1+\mathtt{v}_1
% 	+\sqrt{b} \mathtt{v}_2+\mathtt{z}_1)
% 	-
% 	H_4( \mathtt{v}_1
% 	+\sqrt{b} \mathtt{v}_2+\mathtt{z}_1)
% 	\\  \label{ApEq3ppn}
% 	\mathfrak{r}_{\mathtt{v}_1}
% 	& = &
% 	H_4(\mathtt{v}_1+
% 	\sqrt{b}	\mathtt{v}_2+
% 	\mathtt{z}_1)
% 	-
% 	H_2( \sqrt{b}\mathtt{v}_2+\mathtt{z}_1) \\  \label{ApEq4ppn}
% 	\mathfrak{r}_{\mathtt{v}_2}  & = & H_5(\sqrt{a}	\mathtt{v}_1+\mathtt{v}_2+\mathtt{z}_2)
% 	-H_6( \sqrt{a}\mathtt{v}_1+\mathtt{z}_2).
% \end{eqnarray}
% to be zero.
   	\label{V4Th5p} 
   \end{theorem}	
   \begin{proof}  
See Appendix \ref{AppB}.
\end{proof} 
%
%      Appendix \ref{LIC1} provides examples illustrating the formation of  functionals associated with the active entropy terms, the power constraints, and the normalization constraints.
%      
Note that the condition established in Theorem \ref{V4Th5p} is necessary but not sufficient for global optimality. Determining the global optimum additionally requires identifying the optimal strategy, determining the active constraints, and maximizing the corresponding weighted sum-rate with respect to the power and bandwidth allocation variables.

Note that  \ref{13-0} to \ref{NNEq7} are valid for vector inputs as well.  
Consider  optimum vector random code-books for public and private messages (four core random vectors), each associated with a given covariance matrix.  Compound random vectors  are formed as linear combinations of  the four core random vectors.   
Appendix \ref{LIC2} shows that vector Gaussian distribution is the unique density, among densities with the same covariance matrix, for which the first-order variation of the entropy vanishes.   A given (fixed) set of  covariance matrices for core random vectors result in a fixed set of covariance matrices for compound random vectors.
Fixing covariance matrices, it follows that compound random vectors should be Gaussian. Noting the linear programming structure in \ref{13-0} to \ref{NNEq7}, it turns out that the corresponding maximizing solution can be expressed as a linear combination of entropy terms, corresponding to compound random vectors, appearing in active constraints.  
A vanishing first order variation of the objective function  is  equivalent to vanishing first order variations for all such (active) entropy terms.  

\subsection{Degradedness in 2-users Weak GIC with Gaussian codebooks}   \label{S4.2}

For simplicity of notation, Theorem \ref{V4Th2}, presented next, is established for scalar inputs. It is straightforward to verify that the corresponding result can be generalized to vector inputs. 

\begin{theorem} 
	For Gaussian codebooks, message $U_2$ at $Y_1$ is a degraded version of message $U_2$ at $Y_2$, message $U_1$ at $Y_2$ is a degraded version of message $U_1$ at $Y_1$, 	message $U_1$ at $Y_2$ after decoding of $U_2$ is a degraded version of message $U_1$ at $Y_1$ after decoding of $U_2$, and 
	message $U_2$ at $Y_1$ after decoding of $U_1$ is a degraded version of message $U_2$ at $Y_2$ after decoding of $U_1$.
	\label{V4Th2} 	
\end{theorem}	

\begin{figure}[p]
	\centering
	\includegraphics[width=0.55\textwidth]{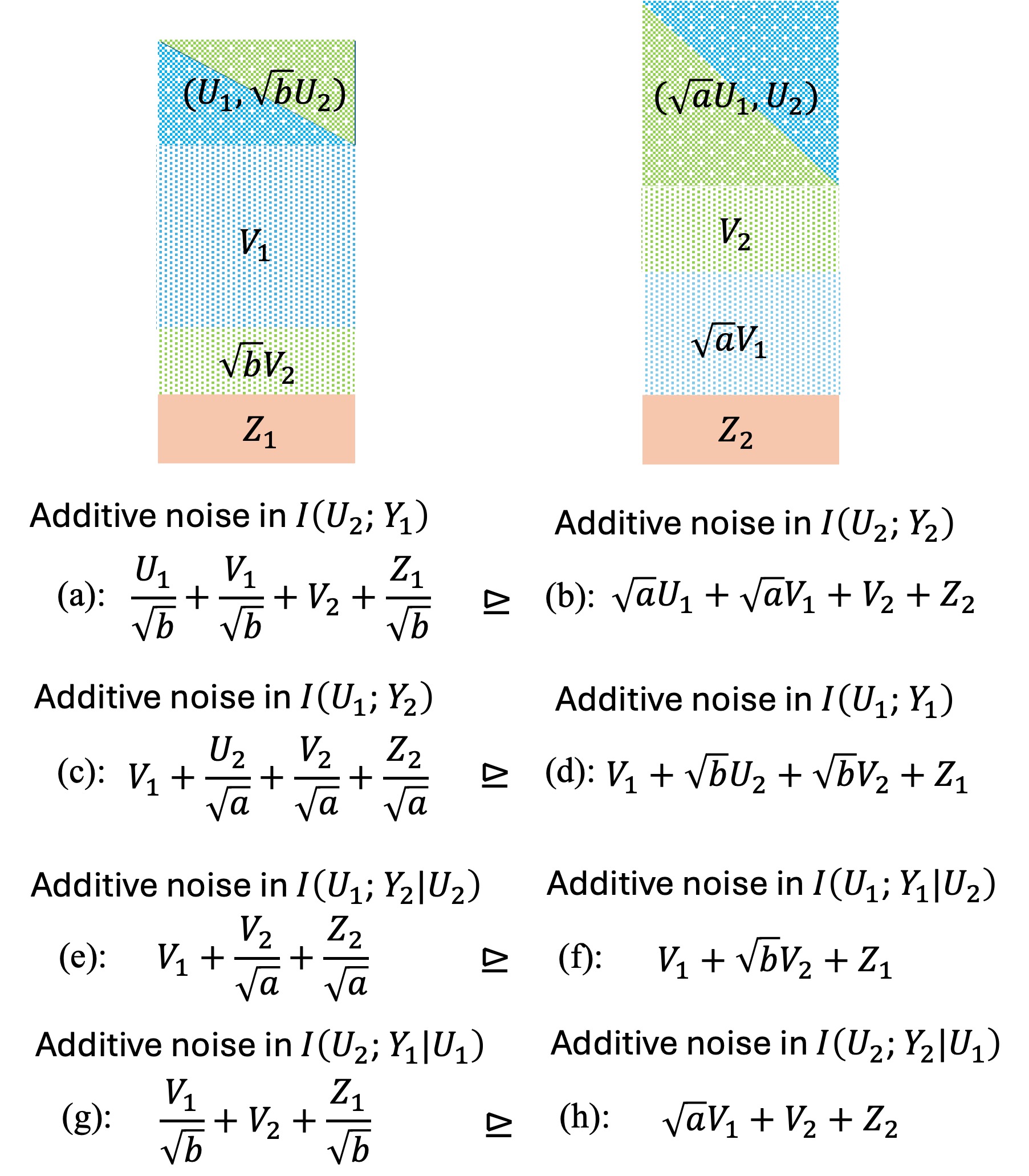}
	\caption{Configurations used in the proof of Theorem \ref{V4Th2}.  Notation $\eth \trianglerighteq \wp$ means capacity of a channel with input $\mathfrak{I}$ and additive noise term $\eth$ (independent of $\mathfrak{I}$) is smaller  as compared to that of a channel with input $\mathfrak{I}$ and additive noise term $\wp$ (independent of $\mathfrak{I}$). These relationships emerge since $a\leq 1$ and $b\leq 1$.}
	\label{Fig-UVZ3N}
\end{figure}

\begin{proof}
	The proof is established  by considering the additive Gaussian noise channels related by $\trianglerighteq$ in Fig.~\ref{Fig-UVZ3N} --- note that $a,b\leq 1$.

\end{proof}

\subsection{Optimality of Gaussian for Single Letter  (Scalar) Compound  Variables} \label{S4.3}

\begin{theorem} \label{TCon0}
	Consider a phase where both users are active. Independent and identically  distributed  Gaussian code-books (single letter) optimize the weighted sum-rate. 
\end{theorem}  
\begin{proof}
	
From  Theorem~\ref{V4Th5p} and Appendix  \ref{LIC2}, it follows that  that the optimum densities for the vectors $\mathtt{u}_1$, $\mathtt{v}_1$, $\mathtt{u}_2$, $\mathtt{v}_2$, $\mathtt{y}_1$ and $\mathtt{y}_2$ will be Gaussian.
Let $\Re_{\mathtt{u}_1}$, $\Re_{\mathtt{v}_1}$, $\Re_{\mathtt{u}_2}$, and $\Re_{\mathtt{v}_2}$ denote the optimum covariance matrices for the Gaussian code-books $\mathtt{u}_1$, $\mathtt{v}_1$, $\mathtt{u}_2$, and $\mathtt{v}_2$, respectively\footnote{Note the degradedness properties established in Theorem~\ref{V4Th2} are valid for vector messages with given covariance matrices.}. 
Let us focus on the case that the rate of $\mathtt{u}_2$  is governed by the successive decoding at $Y_1$, i.e.,
\begin{eqnarray} 	\label{ApEq1p}
	\mathbf{r}_{\mathtt{u}_2}
& = &
	H( \overbrace{\mathtt{u}_1+\mathtt{v}_1
		+\sqrt{b}\mathtt{u}_2+\sqrt{b}\mathtt{v}_2+\mathtt{z}_1}^{\mathtt{s}_{\mathtt{u}_2}})
	-
	H( \overbrace{\mathtt{u}_1
		+\mathtt{v}_1+\sqrt{b}\mathtt{v}_2+\mathtt{z}_1}^{\mathtt{n}_{\mathtt{u}_2}})
	=
	H(\mathtt{s}_{\mathtt{u}_2})
	-
	H(\mathtt{n}_{\mathtt{u}_2}) 
	\\ \label{ApEq2p}
		\mathbf{r}_{\mathtt{u}_1}
& = &
	H( \overbrace{\mathtt{u}_1+\mathtt{v}_1
		+\sqrt{b} \mathtt{v}_2+\mathtt{z}_1}^{\mathtt{s}_{\mathtt{u}_1}})
	-
	H( \overbrace{\mathtt{v}_1
		+\sqrt{b} \mathtt{v}_2+\mathtt{z}_1}^{\mathtt{n}_{\mathtt{u}_1}})
	=
	H(\mathtt{s}_{\mathtt{u}_1})
	-
	H(\mathtt{n}_{\mathtt{u}_1}) \\  \label{ApEq3p}
			\mathbf{r}_{\mathtt{v}_1}
	& = &
	H( \overbrace{\mathtt{v}_1+
		\sqrt{b}	\mathtt{v}_2+
		\mathtt{z}_1}^{\mathtt{s}_{\mathtt{v}_{1}}})
	-
	H( \overbrace{\sqrt{b}\mathtt{v}_2+\mathtt{z}_1}^{\mathtt{n}_{\mathtt{v}_{1}}})
	=
	H(\mathtt{s}_{\mathtt{v}_{1}})
	-
	H(\mathtt{n}_{\mathtt{v}_{1}})  \\ \label{ApEq4p}
			\mathbf{r}_{\mathtt{v}_2}
	& = &
	H( \overbrace{
			\mathtt{v}_2+\sqrt{a}\mathtt{v}_1+
		\mathtt{z}_2}^{\mathtt{s}_{\mathtt{v}_{2}}})
	-
	H( \overbrace{\sqrt{a}\mathtt{v}_1+\mathtt{z}_2}^{\mathtt{n}_{\mathtt{v}_{2}}})
	=
	H(\mathtt{s}_{\mathtt{v}_{2}})
	-
	H(\mathtt{n}_{\mathtt{v}_{2}})  \\ \label{ApEq5p}
		\mathbf{r}_{\mathtt{u}_1}+\mathbf{r}_{\mathtt{u}_2}
	& = &
	H( \overbrace{\sqrt{a}\mathtt{u}_1+\mathtt{u}_2+\sqrt{a}\mathtt{v}_1+
	\mathtt{v}_2+
		\mathtt{z}_2}^{\mathtt{s}_{\mathtt{u}_{1\!,2}}})
	-
	H( \overbrace{\sqrt{a}\mathtt{v}_1+\mathtt{v}_2+\mathtt{z}_2}^{\mathtt{n}_{\mathtt{u}_{1\!,2}}})
	=
	H(\mathtt{s}_{\mathtt{u}_{1\!,2}})
	-
	H(\mathtt{n}_{\mathtt{u}_{1\!,2}})
\end{eqnarray}
where $H(\mathtt{a})$ denotes the entropy of the vector $\mathtt{a}$.  
Note that $\sqrt{b}\mathtt{v}_2+\mathtt{z}_1$  acts as an additive Gaussian noise vector in \ref{ApEq1p}, \ref{ApEq2p}, \ref{ApEq3p}. Consequently,  in the optimizing solution, the rate values in \ref{ApEq1p}, \ref{ApEq2p}, \ref{ApEq3p} are maximized when 
$\mathtt{u}_2$, $\mathtt{u}_1$, $\mathtt{v}_1$ correspond to water filling over the eigenvectors of $\Re_{\mathtt{v}_2}$. 
Likewise, $\mathbf{r}_{\mathtt{v}_2}$ in \ref{ApEq4p} is maximized if $\mathtt{v}_2$ correspond to water filling over the eigenvectors of $\Re_{\mathtt{v}_1}$. These requirements  are  both satisfied if eigenvectors of $\Re_{\mathtt{u}_1}$, $\Re_{\mathtt{v}_1}$, $\Re_{\mathtt{u}_2}$, $\Re_{\mathtt{v}_2}$ are the same, and power values $E(\|\mathtt{u}_1\|^2)$, $E(\|\mathtt{v}_1\|^2)$, $E(\|\mathtt{u}_2\|^2)$, $E(\|\mathtt{v}_2\|^2)$ are uniformly allocated to these eigenvectors. It follows that, such a basis will be the eigenvectors corresponding to signal $\sqrt{a}\mathtt{u}_1+\mathtt{u}_2$ as well as to the additive noise $\sqrt{a}\mathtt{v}_1+\mathtt{v}_2$
in \ref{ApEq5p}.  Since the resulting power allocation for the additive noise channel in \ref{ApEq5p} turns out to be uniform, it follows that  	$\mathbf{r}_{\mathtt{u}_1}+\mathbf{r}_{\mathtt{u}_2}$ in \ref{ApEq5p} is maximized as well.  
This means the weighted sum-rate can be maximized by using the same  eigenbasis, say an identity matrix, for all channels and  relying on uniform power allocation over this basis. Consequently, the optimum densities for $\mathtt{u}_1$, $\mathtt{v}_1$, $\mathtt{u}_2$, $\mathtt{v}_2$, $\mathtt{y}_1$, $\mathtt{y}_2$ will be  i.i.d. Gaussian.
\end{proof}

%In the following, $U_1$, $V_1$, $U_2$, and $V_2$ are referred to as {\em core random variables}, while their linear combinations are referred to as {\em compound random variables}.
% 
\section{\Large Optimality of i.i.d. Single Letter Gaussian Code-books for  $U_1$, $V_1$, $U_2$, $V_2$ } \label{S5}

\subsection{Gaussian Compound Random Variables Result in Gaussian Core Random  Variables}
\label{S5.1}

\begin{figure}[h]
	\centering
	\includegraphics[width=0.65\textwidth]{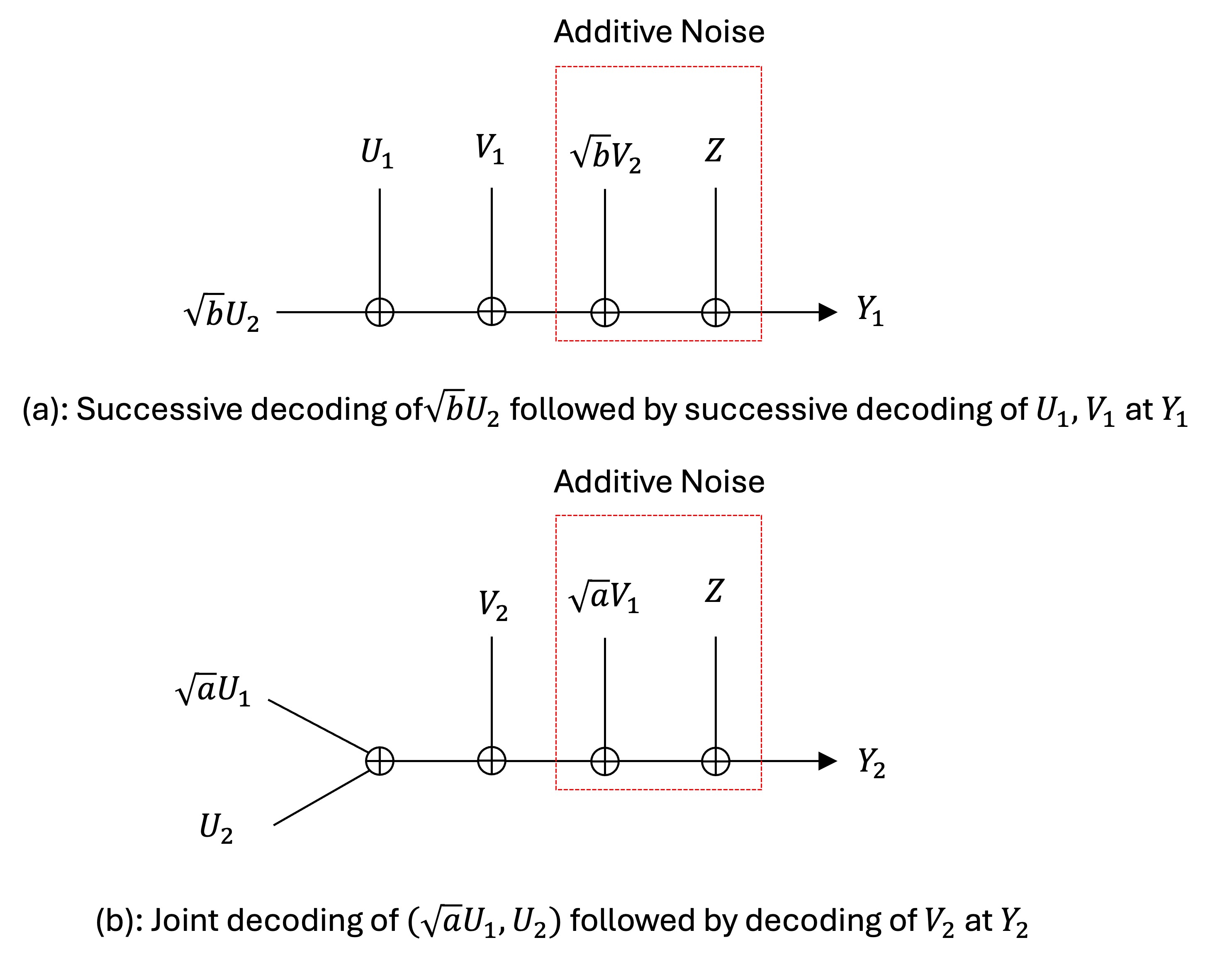}
	\caption{Channel models depicting decoding methods discussed in Theorem \ref{V4Th3} where \ref{F4V4L}(a) corresponds to successive decoding of $\sqrt{b} U_2$, $U_1$ and $V_1$ at $Y_1$,  and \ref{F4V4L}(b) corresponds to joint decoding of $(U_1, \sqrt{b} U_2)$ followed by decoding of $V_1$ at $Y_1$.}
\label{F4V4L}
\end{figure}

Theorem~\ref{V4Th3} establishes that,  $U_1$, $V_1$, $U_2$, $V_2$ are a unique linear combination of compound random variables formed by successive decoding at $Y_1$ or at $Y_2$.
This property shows that if compound random variables are jointly Gaussian, then $U_1$, $V_1$, $U_2$, $V_2$ will be Gaussian as well. 
\begin{theorem}  
	There exits an invertible $4\times 4$ matrix allowing to express core random variables in terms of compound random variables. 
	\label{V4Th3} 	% {V4Th1}
\end{theorem}	
\begin{proof} For Gaussian compound variables, from Section~\ref{S4.2},  we  have the degraded cases depicted in Fig.~\ref{Fig-UVZ3N}. 
	Without loss of generality, let us assume $b\neq 0$, and focus on the case that successive decoding of public message(s) is performed at $Y_1$. Consider compound random variables  $C_1$ to $C_4$ involved in successive decoding at $Y_1$. We have
	\begin{eqnarray} \label{V4E26NN}
		C_1 & = & U_1+V_1+ \sqrt{b} U_2+\sqrt{b} V_2   \\ \label{V4E27NN}
		C_2 & = & U_1+V_1+ \sqrt{b} V_2   \\  \label{V4E28NN}
		C_3 & = & V_1+\sqrt{b}  V_2   \\ \label{V4E29NN} 
		C_4 & = & \sqrt{b}   V_2.
	\end{eqnarray} 
	Matrix of linear coefficients forming \ref{V4E26NN}, \ref{V4E27NN}, \ref{V4E28NN}, \ref{V4E29NN} is equal to
	\begin{eqnarray}  \label{V4E32}
		\begin{bmatrix} 
		1 & 1 & \sqrt{b}  & \sqrt{b}  \\
		1 & 1 &   0  & \sqrt{b}  \\
		0 & 1 & 0  & \sqrt{b} \\
		0 & 0& 0 &  \sqrt{b} 
		\end{bmatrix}. 
	\end{eqnarray}
	It follows that the matrix in \ref{V4E32} is invertible $\forall b\neq 0$.  Figure~\ref{F4V4L} depicts the underlying additive noise channels. 
\end{proof}  
A similar result follows if $U_1$ or $U_2$ is zero.   

\vspace{0.25cm}
\noindent
{\bf Remark 3:} Proof of Theorem \ref{V4Th3} focused on the model in Fig.~\ref{F4V4L}.  
A second case concerns boundary points where only one of the users has a public message. Without loss of generality, let us consider the case that only user 2 has a public message. In this case,  \ref{NNEq6ttV} to  \ref{NNEq7} change to:
\begin{eqnarray} \label{EqN49}
R_{V_1}  & \leq  & I(V_1;Y_1|U_2) \\  \label{EqN50}
R_{V_2}  & \leq & I(V_2;Y_2|U_2) \\  \label{EqN51}
R_{U_2}  & \leq  & I(U_2;Y_2) \\  \label{EN52}
R_{U_2}  & \leq  & I(U_2;Y_1).
\end{eqnarray}
At least three constraints among \ref{EqN49} to \ref{EN52} should be active.  Let us consider the case that \ref{EqN49}, \ref{EqN50} and \ref{EN52} are active.   		    
%\begin{figure}[htbp]
%\centering
%\includegraphics[width=0.7\textwidth]{F4V4F3pp3}
%\caption{Channel models depicting decoding methods corresponding to constraints \ref{EqN49} to \ref{EN52}, where (a),(b) correspond to \ref{EqN49}, \ref{EqN50}, \ref{EqN51} being active,  and (c),(d) correspond to \ref{EqN49}, \ref{EqN50}, \ref{EN52} being active.} 
%\label{F4V4L2}
%\end{figure}
Corresponding additive noise channels involve three compound random variables, namely 
\begin{eqnarray} \label{EqN53}
\hat{C}_1 & = & \sqrt{b}U_2+V_1+\sqrt{b}V_2 \\ \label{EqN54}
\hat{C}_2& = & V_1+\sqrt{b}V_2 \\ \label{EqN55}
\hat{C}_3& = & \sqrt{b}V_2. 
\end{eqnarray} 
This results in the matrix 
\begin{eqnarray}  \label{Eqn56}
\begin{bmatrix} 
	 \sqrt{b} & 1  &  \sqrt{b}  \\
0 &  1 & \sqrt{b}  \\
0& 0 & \sqrt{b}  
\end{bmatrix}.
\end{eqnarray}
Assuming $b\neq 0$,  the matrix in \ref{Eqn56} is full rank.
%
%In addition, the  boundary may include a single point where both users send private messages only.  Such a point will be at the intersection of two segments, one corresponding to (only) user 1 having a public message, and the other one corresponding to (only) user 2 having a public message. It follows that the relevant $2\times 2$ matrices are non-singular. An example is the $2\times 2$ block shown in bold in  \ref{Eqn56}.    
$\blacksquare$

\subsection{Structure of Phases used in Single-letter Time-sharing} \label{S5.2}
%\subsubsection{Preliminaries for Section \ref{S5.2}}
In time-sharing, time axis is divided into multiple non-overlapping segments, called phases hereafter. Each phase uses a fraction of time, a fraction of $P_1$ and a fraction of $P_2$,  to maximize its relative contribution to the cumulative weighted sum-rate.
 Let us focus on a pair of phases. Superscripts $(\,\bar{\cdot}\,)$ and $(\,\Bar{\Bar{\cdot}}\,)$ refer to the first phase and the second phase in the pair.  Power of user 1  allocated to the two phases are denoted as
$\bar{\wp}_{_1}$ and $\Bar{\Bar{\wp}}_{_1}$. Likewise, power of user 2 allocated to the two phases are denoted as $\bar{\wp}_{_2}$ and $\Bar{\Bar{\wp}}_{_2}$.  
%Notations $\mathtt{u}_1, \mathtt{v}_1,\mathtt{u}_2, \mathtt{v}_2$, $\mathtt{x}_1, \mathtt{x}_2$  refer to (vector) code-books and $\mathtt{y}_1, \mathtt{y}_2$ to corresponding outputs. 	Components of a vector are indexed using a superscript, e.g., components of $\mathtt{y}_1$ are denoted as  $y^i_1$.  
Corresponding rate values are denoted as $\bar{\mathbi{R}}_1=\bar{\mathbf{r}}_{\mathtt{u}_1}+\bar{\mathbf{r}}_{\mathtt{v}_1}$, $\bar{\mathbi{R}}_2=\bar{\mathbf{r}}_{\mathtt{u}_2}+\bar{\mathbf{r}}_{\mathtt{v}_2}$ and 
$\Bar{\Bar{\mathbi{R}}}_1=\Bar{\Bar{\mathbf{r}}}_{\mathtt{u}_1}+\Bar{\Bar{\mathbf{r}}}_{\mathtt{v}_1}$, 	 $\Bar{\Bar{\mathbi{R}}}_2=\Bar{\Bar{\mathbf{r}}}_{\mathtt{u}_2}+\Bar{\Bar{\mathbf{r}}}_{\mathtt{v}_2}$. 
%Correlation matrices of a vector $\mathtt{a}$ is denoted as $\Re_\mathtt{a}=E(\mathtt{a}\mathtt{a}^t)$.

%\subsubsection{Main Results of Section \ref{S5.2}}
\begin{theorem} \label{Th11}
	Consider two phases over which both users are active. An optimum solution exists for which the two phases can be merged into one.  
\end{theorem}	

\begin{proof} 
Without loss of generality, let us consider a portion of Phase 1 and a portion of Phase 2 having equal duration. It suffices to show that these two sub-phases can be merged.
Consider maximizing the weighted sum-rate
$\bar{\mathbi{R}}_1+\mu\bar{\mathbi{R}}_2+\Bar{\Bar{\mathbi{R}}}_1+\mu\Bar{\Bar{\mathbi{R}}}_2$
subject to the total power constraints
$\bar{\wp}{1}+\Bar{\Bar{\wp}}{1}$ for user 1 and
$\bar{\wp}{2}+\Bar{\Bar{\wp}}{2}$ for user 2. 
For $\mu<1$, from Theorem~\ref{V4Th2}, $U_2$  is decoded first at $Y_1$ (see Figs.~\ref{Fig-UVZ3N} and \ref{F4V4L}). The decodability constraint imposed by this first layer in the successive decoding governs the rate of $U_2$. Since the codebooks are Gaussian, it follows that water-filling the total power of $U_2$ over the entire bandwidth increases the rate of $U_2$. Upon decoding and removing $U_2$, a similar argument applies to  decoding of $U_1$   at $Y_1$, likewise to all layers involved in successive decoding. It follows that the power values $\bar{\wp}{1}+\Bar{\Bar{\wp}}{1}$ and $\bar{\wp}{2}+\Bar{\Bar{\wp}}{2}$ should be allocated uniformly across the entire time–frequency resource for each message $U_1$, $V_1$, $U_2$, and $V_2$.
\end{proof}

\begin{theorem} \label{Th12p}
	Assume the optimum solution includes a phase where both users are active. 
	There is at most one additional phase over which a single user is active. 
\end{theorem}	  
\begin{proof} 
	Merging of phases and water-filling over the resulting overall band, presented in Theorem~\ref{Th11},  can be  applied until (potentially) a phase with a single user emerges. \end{proof}

\vspace{0.25cm}
\noindent
{\bf Remark 4:} 
Theorems~\ref{Th11} and \ref{Th12p} can be applied recursively to merge any number of phases in which both users are active into a single two-user phase. A phase occupied by only one user, if present, reduces to a point-to-point Gaussian channel, whose capacity is achieved by a single-letter Gaussian codebook.
It follows that any point on the upper concave envelope can be attained by time-sharing between at most two single-letter Gaussian codebooks: one associated with a two-user phase and, if present, one associated with a single-user phase.
$\blacksquare$

\begin{theorem} \label{Th8N}
Embedding information in the selection of strategy 
can not improve the  weighted sum-rate. 
\end{theorem}	  
\begin{proof} 
	Let  us define selection alphabets $\eth_1\in\{1,2\}$ and  $\eth_2\in\{1,2\}$, for user 1 and  user 2,  in selecting the first phase and the second phase, respectively. 
	Let us focus on the case that user 2 occupies the  single user phase, i.e.,  $\eth_1=1$ and $\eth_2\in\{1,2\}$.  To maximize $R_2$, user 2 can optimize: (1)  the power allocation across the two phases, and (2) embedding information in  the selection of $\eth_2\in\{1,2\}$.  Since code-books are Gaussian, a uniform height of power across the two phases/bands (water filling) maximizes the rate of user 2. The reason is that any excess power in the second phase can spill into the first phase, and create a Gaussian layer of 
	For such a balanced power allocation,  the density of $\mathtt{y}_2$ will be i.i.d. Gaussian regardless of the value of $\eth_2$. This means $\mathtt{y}_2$ is independent of $\eth_2$, hence selection of $\eth_2$ cannot convey any information. 
\end{proof} 

\subsection{Single-letter Code-books are Zero-mean}  \label{S5.3}

Since power constraints are forced to be satisfied with equality, a stationary solution may include cases that code-books' densities have a non-zero statistical mean. Following example clarifies this point.  

\vspace{0.25cm}
\noindent
{\bf Example:}
\begin{figure}[h]
	\centering
	\includegraphics[width=0.6\textwidth]{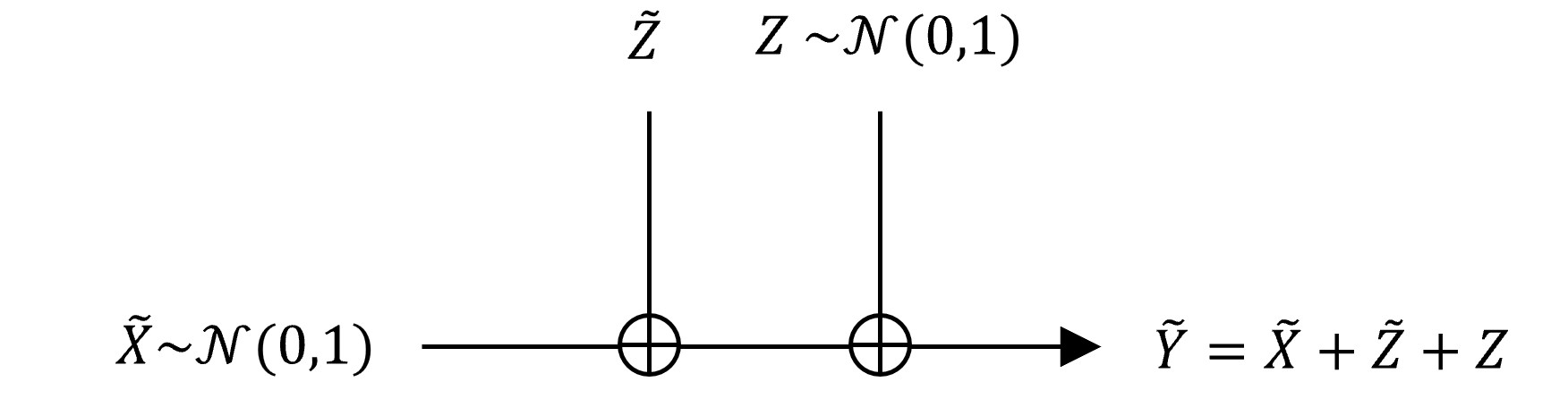}
	\caption{Example of a channel where the stationary solution for mutual information may result in a maximum or a minimum, according to the statistical mean of $\tilde{Z}$.}
	\label{Example}
\end{figure}
Consider the channel in Fig.~\ref{Example}, where $\tilde{X}$, $\tilde{Z}$ and $Z$ are independent, and 
$ \int \vartheta^2f_{\tilde{Z}}(\vartheta)d\vartheta  =  P_{\!\tilde{Z}}$. 
Let us define
\begin{eqnarray}
	\hat{f} & = & f_{\tilde{X}}\ast f_{\tilde{Z}}\ast {\cal N}(0,1)\\
	&=&f_{\tilde{Z}}+{\cal N}(0,2) \\
	\check{f} & = & f_{\tilde{Z}}\ast {\cal N}(0,1)
\end{eqnarray}
where ${\cal N}(\mathfrak{u},\mathfrak{s})$  is a Gaussian density with statistical average $\mathfrak{u}$ and variance $\mathfrak{s}$. 
We have
\begin{equation}
	I(\tilde{X};\tilde{Y})=	\mathsf{H}^{\hat{f}}-\mathsf{H}^{\check{f}}
	\label{New48}
\end{equation}
where $\mathsf{H}^{\hat{f}}$ and $\mathsf{H}^{\check{f}}$ are entropy values for densities $\hat{f}$ and $\check{f}$. 
It follows that 
\begin{eqnarray}  
	\min_{\hat{f},\check{f}}I(\tilde{X};\tilde{Y}) ~~\mbox{is achieved for}  &  \!\!\!\!\!\!\!\!\!\!f_{\tilde{Z}}={\cal N}(0,P_{\!\tilde{Z}}) \\
	\max_{\hat{f},\check{f}}I(\tilde{X};\tilde{Y}) ~~\mbox{is achieved for}  & \!\!f_{\tilde{Z}}={\cal N}(\sqrt{P_{\!\tilde{Z}}},0).
\end{eqnarray}  
A non-zero statistical mean entails the power $P_{\!\tilde{Z}}$ is intentionally wasted to avoid interference.  
$\blacksquare$

\vspace{0.15cm} 
Theorem~\ref{New1} establishes that the codebook densities corresponding to $U_1$, $V_1$, $U_2$, and $V_2$ are zero-mean. Consequently, all optimization problems can be formulated, without loss of optimality, in terms of zero-mean densities.

\begin{theorem}
	Code-books' densities for (single letter) $U_1$, $V_1$, $U_2$, $V_2$ are zero mean for $\forall P_1, \forall P_2$. 
	\label{New1} 
	\end{theorem}	
	\begin{proof}  
	For $\mu<1$, the codebook densities corresponding to  $U_1$ and $U_2$ are zero-mean. The reason is that, instead of wasting part of the power budgets $P_{U_1}$ and/or $P_{U_2}$ on non-zero mean values, the same power can be used to increase the variances of the corresponding codebooks. This, in turn, increases $R_{U_1}$ and/or $R_{U_2}$ while preserving the requirement that the public messages be decodable at both receivers. On the other hand, if the codebook densities corresponding to the private messages $V_1$ and/or $V_2$ have non-zero means, the power allocated to those means can instead be reassigned to the corresponding public messages. This strictly increases $R_{U_1}$ and/or $R_{U_2}$ while maintaining the decodability of both the public and private messages.
	\end{proof}  

It is of interest to derive explicit expressions and identify the conditions governing the formation of the various segments of the boundary. This task is challenging due to the large number of parameters involved and the intricate interplay among the corresponding conditions. Section~\ref{ABC1} presents a method that significantly simplifies these derivations. In addition, it shows how the boundary can be traversed  in a continuous manner. 

\section{Covering the Boundary using Power Reallocation Steps}
\label{ABC1}

The capacity region in the single-letter case is covered by starting from the point that maximizes $R_1$ and proceeding counterclockwise along the lower boundary, corresponding to $\mu<1$. In a sequence of infinitesimal steps, $R_2$ is gradually increased in exchange for a decrease in $R_1$.
Each step involves infinitesimal reallocations of power among the messages. The reallocated power amounts, denoted by $\delta P_1$ and $\delta P_2$, are chosen sufficiently small so that the {\em coding strategy} remains unchanged within the step; any change in strategy can occur only at the beginning of a subsequent step.

Consider an infinitesimal step from a starting point, denoted by superscript $s$, to an ending point, denoted by superscript $e$. The slope $\mathbf{\Upsilon}$ of such a step is defined as
\begin{equation} 	
	\mathbf{\Upsilon}= \frac{\Delta\!R_2} {\Delta\!R_1} =  \frac{R^e_{V_2}+R^e_{U_2}-R^s_{V_2}-R^s_{U_2}}{
		R^s_{V_1}+R^s_{U_1}-R^e_{V_1}-R^e_{U_1} } \triangleq \frac{\mathbf{N}}{\mathbf{D}} 
	\label{Eq3}
\end{equation} 
where $(R^s_{U_1},R^s_{V_1})$ and $(R^s_{U_2},R^s_{V_2})$ denote the public and private rates of users 1 and 2, respectively, at the starting point. Likewise, $(R^e_{U_1},R^e_{V_1})$ and $(R^e_{U_2},R^e_{V_2})$ denote the corresponding public and private rates at the end point.
Note that $\Delta R_1$ and $\Delta R_2$ are defined to be positive quantities. In particular, $\Delta R_1$ is defined as the rate $R_1$ at the starting point minus the rate $R_1$ at the end point. Starting from a point on the boundary, 
the optimality of the next point is characterized by: 
\begin{equation}
	\mbox{Maximize}~~\mathbf{\Gamma}
	=
	\sqrt{(\Delta R_1)^2+(\Delta R_2)^2}~~\mbox{for different values of}~~ \mathbf{\Upsilon}.
	\label{Eq4}
\end{equation}
 \begin{figure}[h]
	\centering
	\includegraphics[width=0.55\textwidth]{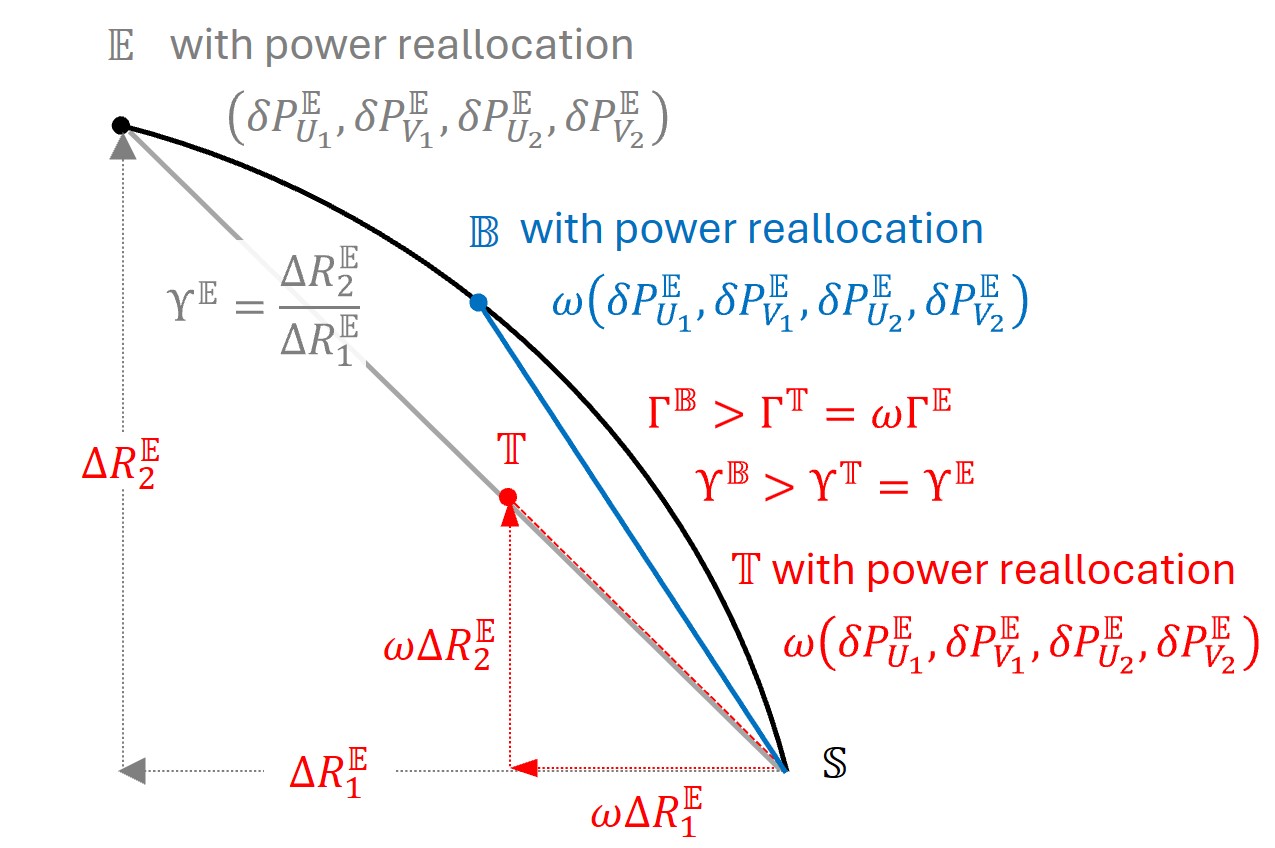}
	\caption{A step along the boundary: $\mathbf{\Upsilon}$ and $\mathbf{\Gamma}$ are  expressed as functions of  $\omega$ which governs how one can  move from $\mathbb{S}$ to  $\mathbb{E}$ through an intermediate point $\mathbb{B}$}. 
	\label{FigTime2}
\end{figure}  
Figure~\ref{FigTime2} depicts a segment on the boundary from a starting point $\mathbb{S}$ to an end point $\mathbb{E}$.  With some misuse of notation, superscripts are used to refer to points inside or on the boundary. 	
Assume power reallocation vector for point  $\mathbb{E}$ is equal to 
$({\delta}P^\mathbb{E}_{U_1},{\delta} P^\mathbb{E}_{V_1},{\delta}P^\mathbb{E}_{U_2},{\delta}P^\mathbb{E}_{V_2})$. Consider the line connecting points $\mathbb{S}$ and $\mathbb{E}$ with a time interpolation factor $\omega\in[0,1]$ where $\omega=0$ and $\omega=1$ correspond to points $\mathbb{S}$ and $\mathbb{E}$, respectively.
Time interpolation achieves point $\mathbb{T}$ inside the capacity region, corresponding to an effective (interpolated) power reallocation vector 
$\omega({\delta} P^\mathbb{E}_{U_1},{\delta}^\mathbb{E} P_{V_1},{\delta}P^\mathbb{E}_{U_2},{\delta} P^\mathbb{E}_{V_2})$. 
Consider the power reallocation vector 
$\omega({\delta} P^\mathbb{E}_{U_1},{\delta}^\mathbb{E} P_{V_1},{\delta}P^\mathbb{E}_{U_2},{\delta} P^\mathbb{E}_{V_2})$ with optimum code-books' densities,  
resulting in the point $\mathbb{B}$ on the boundary corresponding to  $\mathbf{\Upsilon}^{\mathbb{B}}(\omega)$, $\mathbf{\Gamma}^{\mathbb{B}}(\omega)$.  
Relying on  a simple time interpolation to achieve  point $\mathbb{T}$ and optimum code-books' densities  to achieve  points $\mathbb{E}$ and $\mathbb{B}$, we have 
\begin{eqnarray} 
	\mathbf{\Upsilon}^{\mathbb{B}}  & > & \mathbf{\Upsilon}^{\mathbb{T}} ~=~ \mathbf{\Upsilon}^{\mathbb{E}} \\
	\mathbf{\Gamma}^{\mathbb{B}} & > & \mathbf{\Gamma}^{\mathbb{T}}   ~=~    \omega \mathbf{\Gamma}^{\mathbb{E}}.
\end{eqnarray}

Next, the condition for a power reallocation vector to be {\em boundary achieving}  is discussed. 
Let us consider a step along the boundary which is small enough such that the coding strategy remains the same within the step. Let us assume  $(\hat{\Delta}\!P_1,\hat{\Delta}\!P_2)$ is the power reallocation vector corresponding to an end point beyond which a change in strategy is needed, and consider 
\begin{equation}
	(\Delta\!P_1,\Delta\!P_2): \Delta\!P_1\leq \hat{\Delta}\!P_1~~\mbox{and}~~\Delta\!P_2\leq \hat{\Delta}\!P_2.
	\label{E57NN} 
\end{equation}
Let us define $\upsilon\leq \mu^s$, where $\mu^s$ is the value of $\mu$ at the starting point, and the set 
$\Bar{\mathsf{S}}_\upsilon$ as
\begin{eqnarray} \label{Ae7}
	\Bar{\mathsf{S}}_\upsilon  =  \left\{f_{U_1},f_{V_1},f_{U_2},f_{V_2}\!:\! ~\mbox{outgoing slope at the starting point is}~\upsilon \triangleq  \min_{(\delta P_1,\delta P_2)\in [0,\Delta\!P_1]\times [0,\Delta\! P_2]}\mathbf{\Upsilon}\right\}.
\end{eqnarray}
Set $\Bar{\mathsf{S}}_\upsilon$ is defined over all possible code-books' densities, including Gaussian.
Each member of \ref{Ae7} corresponds to a power reallocation vector $(\delta P_1,\delta P_2)\in [0,\Delta\!P_1]\times [0,\Delta\! P_2]$. This correspondence is potentially many-to-one since multiple choices for densities $(f_{U_1},f_{V_1},f_{U_2},f_{V_2})$, with the same $(\delta P_1,\delta P_2)$, may achieve the same $\mathbf{\Upsilon}=\upsilon$.  Given $\mathbf{\Upsilon}=\upsilon$, the size of the set $\Bar{\mathsf{S}}_\upsilon$ is reduced by limiting it to choice(s) which maximize $\mathbf{\Gamma}$. 
Maximum value  of $\mathbf{\Gamma}$ over the set $\Bar{\mathsf{S}}_\upsilon$ is  denoted as $\varkappa_\upsilon$. Let us consider a second set $\Bar{\Bar{\mathsf{S}}}_{\upsilon}$ where 
\begin{equation}
	\Bar{\Bar{\mathsf{S}}}_{\upsilon}\subseteq 
	\Bar{\mathsf{S}}_\upsilon:~\mathbf{\Gamma}=\varkappa_\upsilon.
	\label{Eq58New65}
\end{equation}
The set $\Bar{\Bar{\mathsf{S}}}_{\upsilon}$ includes a point on the boundary with 
\begin{equation}
	\mathbf{\Upsilon}=\upsilon~~\mbox{and}~~\mathbf{\Gamma}=\varkappa_\upsilon\triangleq \max_{\mathbf{\Upsilon}=\upsilon} \mathbf{\Gamma}.
	\label{Eq58New}
\end{equation}
We are interested in establishing that the size of $\Bar{\Bar{\mathsf{S}}}_{\upsilon}$ can be reduced, by increasing $\upsilon$, such that the shrunken set includes a single element, say $\zeta$. Since $\Bar{\Bar{\mathsf{S}}}_{\upsilon}$ always includes a point on the boundary, it  follows that $\zeta$ falls on the boundary.  In addition, we need to show that the rest of the boundary can be covered starting from $\zeta$. Theorem \ref{V4Th6} addresses these requirements.

\begin{theorem} 
	Cardinality of the set $\Bar{\Bar{\mathsf{S}}}_{\upsilon}$ can be reduced, by increasing $\upsilon<\mu^s$, in a recursive manner, such that the final set is associated with a single $(\delta P_1,\delta P_2)$. 
	\label{V4Th6}	
	\end{theorem}	
	\begin{proof} 
	See Appendix \ref{PrX13}.
		\end{proof} 
		
Referring to Theorem \ref{V4Th6}, using $(\breve{\delta} P_1,\breve{\delta} P_2)$ instead of $(\delta^m P_1,\delta^m P_2)$, $m=1,\ldots,M$ (see expression \ref{Ae14}) is accompanied by a movement in clockwise direction, i.e., reaching from  $(\upsilon, \gamma)$ to $(\breve{\upsilon},	\breve{\gamma})$, where 
	\begin{equation}
		(\upsilon, \varkappa_{\upsilon})  \leadsto  (\breve{\upsilon},\varkappa_{\breve{\upsilon}}):~~ 
		\breve{\upsilon}  >   \upsilon~~\mbox{and}~~\varkappa_{\breve{\upsilon}}<\varkappa_{\upsilon}.
		\label{Eq53V4}	
	\end{equation}
	Such a movement can continue in a recursive manner until the step size is small enough to include a single power reallocation vector, i.e., 
	\begin{equation}
		\exists i\in[1,\ldots,M]:~(\breve{\delta} P_1,\breve{\delta} P_2)=(\delta^i P_1,\delta^i P_2)
		\label{Eq64}	
	\end{equation}
	with the resulting $(\breve{\delta} P_1,\breve{\delta} P_2)$ achieving to a unique point on the boundary. 
	Theorem \ref{V4Th6} entails, relying on Pareto minimal power reallocation, the past history in moving counterclockwise along the boundary is captured solely by the starting point in each step.

\vspace{0.25cm}
\noindent
{\bf Remark 5:}     	 
The optimum Pareto minimal power reallocation vector is not unique. However, the corresponding set has a nested structure, and relying on any member of the set will be associated with a unique set of Gaussian code-books.  Different  members of the set of Pareto minimal power reallocation pairs  correspond to different step sizes. This property allows covering the boundary in a continuous manner.   
To clarify this point, let us consider two nested Pareto minimal power reallocation vectors $(\dot{\delta} P_1, \dot{\delta} P_2)$ and $(\ddot{\delta} P_1, \ddot{\delta} P_2)$, where 
\begin{equation}
	\dot{\delta} P_1 \leq \ddot{\delta} P_1~~\mbox{and}~~\dot{\delta} P_2 \leq \ddot{\delta} P_2.
	\label{EQNNN1}
\end{equation}
These power reallocation vectors, in conjunction with Gaussian code-books, achieve two successive points on the boundary, namely 
\begin{equation}
	(\mathbf{\Upsilon}_{\!1},\mathbf{\Gamma}_{\!1})=(\dot{\upsilon},\varkappa_{\dot{\upsilon}})~~\mbox{and}~~(\mathbf{\Upsilon}_{\!2},\mathbf{\Gamma}_{\!2})=(\ddot{\upsilon},
	\varkappa_{\ddot{\upsilon}})
\end{equation}
satisfying $	\ddot{\upsilon}\leq \dot{\upsilon}~~\mbox{and}~~\varkappa_{\ddot{\upsilon}}\geq  
\varkappa_{\dot{\upsilon}}.$~~~
$\blacksquare$

	\vspace{0.15cm}
\section{Converse Results} \label{sec4p}
Consider a phase where both users are active, together with a given power allocation over messages, and given codebooks' densities. 
 Without loss of generality, let us consider the case that $U_2$ is first decoded at $Y_1$, followed by successive decoding of $U_1$ and $V_1$, we have 
\begin{eqnarray} \label{RE1V}\	\mathbf{r}_{\mathtt{u}_1}+	\mathbf{r}_{\mathtt{u}_2}  & = & I(\mathtt{u}_1,\mathtt{u}_2;\mathtt{y}_1)=
	\overbrace{I(\mathtt{u}_2;\mathtt{y}_1) +I(\mathtt{u}_1;\mathtt{y}_1|\mathtt{u}_2)}^{\mbox{\scriptsize Successive Decoding}}=\overbrace{I(\mathtt{u}_1,\mathtt{u}_2;\mathtt{y}_2)}^{\mbox{\scriptsize Joint Decoding}} \\ \label{RE2V}
		\mathbf{r}_{\mathtt{u}_2} & = &  I(\mathtt{u}_2;\mathtt{y}_1)  \\ \label{RE3V}
	\mathbf{r}_{\mathtt{u}_1} & = &  I(\mathtt{u}_1;\mathtt{y}_1|\mathtt{u}_2)  \\ 
  \label{RE4V}
  	\mathbf{r}_{\mathtt{v}_1} & = &  I(\mathtt{v}_1;\mathtt{y}_1|\mathtt{u}_1,\mathtt{u}_2)  \\ \label{RE5V}
  \mathbf{r}_{\mathtt{v}_2} & = &  I(\mathtt{v}_2;\mathtt{y}_2|\mathtt{u}_1,\mathtt{u}_2). 
\end{eqnarray}
\begin{theorem} \label{TCon}
	If probability of error in recovering 
	$\mathtt{u}_1,\mathtt{u}_2,\mathtt{v}_1$ at $\mathtt{y}_1$ and  
	$\mathtt{u}_1,\mathtt{u}_2,\mathtt{v}_2$ at $\mathtt{y}_2$ tend to zero as $\mathtt{t}\rightarrow \infty$, then the rate vector
	$(\mathbi{R}_1=\mathbf{r}_{\mathtt{u}_1}+\mathbf{r}_{\mathtt{v}_1},\mathbi{R}_2=\mathbf{r}_{\mathtt{u}_2}+\mathbf{r}_{\mathtt{v}_2})$ should fall within the optimum  region with independent and identically distributed  single letter Gaussian code-books.  
\end{theorem}

\begin{proof}
	Let us consider the channel models in Fig.~\ref{F4V4L} in conjunction with vectors $\mathtt{u}_1,\mathtt{v}_1,\mathtt{u}_2,\mathtt{v}_2$. Let us assume the set of rates  $(\check{\mathbf{r}}_{\mathtt{u}_1},\check{\mathbf{r}}_{\mathtt{u}_2},\check{\mathbf{r}}_{\mathtt{v}_1},\check{\mathbf{r}}_{\mathtt{v}_2})$. For the decoding strategies in \ref{RE1V} to \ref{RE5V}, let us consider the following cases 
\begin{eqnarray} \label{E84}
	\mbox{Case 1:~~~} 
	\check{\mathbf{r}}_{\mathtt{u}_1}\geq \mathbf{r}_{\mathtt{u}_1},\,\check{\mathbf{r}}_{\mathtt{u}_2}>\mathbf{r}_{\mathtt{u}_2},\,\check{\mathbf{r}}_{\mathtt{v}_1}\geq\mathbf{r}_{\mathtt{v}_1},\,\check{\mathbf{r}}_{\mathtt{v}_2}\geq\mathbf{r}_{\mathtt{v}_2} & \Longrightarrow & P^{\,e}_{\mathtt{u}_1}> 0,\, P^{\,e}_{\mathtt{u}_2}> 0,\, P^{\,e}_{\mathtt{v}_1}> 0,\, P^{\,e}_{\mathtt{v}_2}> 0\\  \label{E85}
		\mbox{Case 2:~~~} 
	\check{\mathbf{r}}_{\mathtt{u}_1}>\mathbf{r}_{\mathtt{u}_1},\,\check{\mathbf{r}}_{\mathtt{u}_2}\geq \mathbf{r}_{\mathtt{u}_2},\,\check{\mathbf{r}}_{\mathtt{v}_1}\geq \mathbf{r}_{\mathtt{v}_1},\,\check{\mathbf{r}}_{\mathtt{v}_2}\geq \mathbf{r}_{\mathtt{v}_2} & \Longrightarrow & P^{\,e}_{\mathtt{u}_1}>\, 0, P^{\,e}_{\mathtt{u}_2}>\, 0, P^{\,e}_{\mathtt{v}_1}>\, 0, P^{\,e}_{\mathtt{v}_2}>\, 0\\  \label{E86}
		\mbox{Case 3:~~~} 
	\check{\mathbf{r}}_{\mathtt{u}_1}\geq \mathbf{r}_{\mathtt{u}_1},\,\check{\mathbf{r}}_{\mathtt{u}_2}\geq \mathbf{r}_{\mathtt{u}_2},\,\check{\mathbf{r}}_{\mathtt{v}_1}> \mathbf{r}_{\mathtt{v}_1},\,\check{\mathbf{r}}_{\mathtt{v}_2}\geq \mathbf{r}_{\mathtt{v}_2} & \Longrightarrow &P^{\,e}_{\mathtt{u}_1}\geq  0,\, P^{\,e}_{\mathtt{u}_2}\geq 0,\, P^{\,e}_{\mathtt{v}_1}> 0,\, P^{\,e}_{\mathtt{v}_2}\geq 0\\  \label{E87}
				\mbox{Case 4:~~~} 
	\check{\mathbf{r}}_{\mathtt{u}_1}\geq \mathbf{r}_{\mathtt{u}_1},\,\check{\mathbf{r}}_{\mathtt{u}_2}\geq \mathbf{r}_{\mathtt{u}_2},\,\check{\mathbf{r}}_{\mathtt{v}_1}\geq  \mathbf{r}_{\mathtt{v}_1},\,\check{\mathbf{r}}_{\mathtt{v}_2}> \mathbf{r}_{\mathtt{v}_2} & \Longrightarrow &
	P^{\,e}_{\mathtt{u}_1}\geq  0,\, P^{\,e}_{\mathtt{u}_2}\geq 0,\, P^{\,e}_{\mathtt{v}_1}\geq 0, \, P^{\,e}_{\mathtt{v}_2}> 0.
\end{eqnarray}
Expressions \ref{E84} and \ref{E85} reflect the fact that in case 1 and case 2, public messages $(U_1,U_2)$ can be recovered/removed neither at $Y_1$ nor at $Y_2$. This causes error propagation to the recovery of $V_1$ at $Y_1$ and $V_2$ at $Y_2$. If $(U_1,U_2)$ can be recovered/removed at both receivers, then cases 3 and 4 can be expressed as 
\begin{eqnarray} 	\label{E88}
	\check{\mathbf{r}}_{\mathtt{u}_1}= \mathbf{r}_{\mathtt{u}_1},\,\,\check{\mathbf{r}}_{\mathtt{u}_2}= \mathbf{r}_{\mathtt{u}_2},\,\,\check{\mathbf{r}}_{\mathtt{v}_1}> \mathbf{r}_{\mathtt{v}_1},\,\,\check{\mathbf{r}}_{\mathtt{v}_2}\geq  \mathbf{r}_{\mathtt{v}_2} & \Longrightarrow &
	P^{\,e}_{\mathtt{u}_1}=  0,\, P^{\,e}_{\mathtt{u}_2}=0,\, P^{\,e}_{\mathtt{v}_1}> 0, \, P^{\,e}_{\mathtt{v}_2}\geq  0.\\ 	\label{E89}
	\check{\mathbf{r}}_{\mathtt{u}_1}= \mathbf{r}_{\mathtt{u}_1},\,\,\check{\mathbf{r}}_{\mathtt{u}_2}= \mathbf{r}_{\mathtt{u}_2},\,\,\check{\mathbf{r}}_{\mathtt{v}_1}\geq  \mathbf{r}_{\mathtt{v}_1},\,\,\check{\mathbf{r}}_{\mathtt{v}_2}> \mathbf{r}_{\mathtt{v}_2} & \Longrightarrow &
	P^{\,e}_{\mathtt{u}_1}=  0,\, P^{\,e}_{\mathtt{u}_2}= 0,\, P^{\,e}_{\mathtt{v}_1}\geq 0, \, P^{\,e}_{\mathtt{v}_2}>  0.
\end{eqnarray}
Above arguments establish that if any rate in \ref{RE1V} to \ref{RE5V} exceeds its corresponding mutual information bound, then, for the decoding strategy given in \ref{RE1V} to \ref{RE5V}, it would be impossible to achieve 
\begin{equation}
P^{\,e}_{\mathtt{u}_1}=  0,~~P^{\,e}_{\mathtt{v}_1}=  0,~~P^{\,e}_{\mathtt{u}_2}=  0,~~P^{\,e}_{\mathtt{v}_2}=  0.
\label{E90}
\end{equation}
	The final step in the proof follows noting that, for any given $\mu$, the region based on \ref{RE1V} to \ref{RE5V} maximizes
\begin{equation}
		\mathbf{r}_1+\mu 	\mathbf{r}_2=
	\mathbf{r}_{\mathtt{u}_1}+\mathbf{r}_{\mathtt{v}_1}+\mu (\mathbf{r}_{\mathtt{u}_2}+\mathbf{r}_{\mathtt{v}_2})
	\label{E91}
	\end{equation}
	 by using independent and identically distributed  single letter Gaussian code-books, while optimizing the corresponding weighted sum-rate over  bandwidth allocation (between phases) and power allocation (between messages/phases). This means in such an optimally enlarged region, if any of the rates exceed the mutual information terms on the right hand sides of   \ref{RE1V} to \ref{RE5V}, the error probability for $(\mathtt{u}_1,\mathtt{v}_1)$ and/or $(\mathtt{u}_2,\mathtt{v}_2)$ will be bounded away from zero. 

\end{proof}

Next, it will be shown that the Han-Kobayashi (HK) achievable rate region, after potentially restricting its feasible set through the imposition of additional but consistent constraints, attains the boundary of the capacity region.

\section{Optimality of the HK Region with Gaussian Code-books} \label{sec4}
Let us consider the 
Expanded Han-Kobayashi constraints\footnote{See expressions 3.2 to 3.15 on page 51 of~\cite{GIC4}, with the changes ($\mbox{current~article} \leftrightarrow [2]$):
	$U_1 \leftrightarrow W_1$, 
	$U_2 \leftrightarrow W_2$,
	$V_1 \leftrightarrow U_1$, 
	$V_2 \leftrightarrow U_2$,
	$R_{U_1}\leftrightarrow T_1$, $R_{U_2}\leftrightarrow T_2$, 
	$R_{V_1}\leftrightarrow S_1$, $R_{V_2}\leftrightarrow S_2$.}
	can be expressed as~\cite{GIC4},
\begin{align} 
	\label{HK1p}
	\mbox{Maximize:}~ & R_1+\mu R_2=R_{U_1}+R_{V_1}+\mu R_{U_2}+\mu R_{V_2}~\mbox{where} \\ \label{HK1}
%	\mbox{Subject to:}~~~ &   \nonumber \\ \label{HK1}
	R_{U_1}    & ~~{\le}~~   I(U_1;Y_1|U_2,V_1)     \\ \label{HK2}
	R_{U_1}  & ~~{\le}~~   I(U_1;Y_2|U_2,V_2)    \\ \label{HK3}
	R_{U_2}   & ~~{\le}~~    I(U_2;Y_1|U_1,V_1)    \\ \label{HK4}
	R_{U_2}   & ~~{\le}~~   I(U_2;Y_2|U_1,V_2)    \\  \label{HK5}
	R_{V_1}  & ~~{\le}~~  I(V_1;Y_1|U_1,U_2)   \\ \label{HK6}
	R_{V_2}  & ~~{\le}~~   I(V_2;Y_2|U_1,U_2)    \\ \label{HK7}  
	R_{U_1}+R_{U_2}   & ~~{\le}~~ I(U_1,U_2;Y_1|V_1)    \\  \label{HK8}
	R_{U_1}+R_{U_2}   & ~~{\le}~~   I(U_1,U_2;Y_2|V_2)  \\  \label{HK9}
	R_{U_1}+R_{V_1}  & ~~{\le}~~  I(U_1,V_1;Y_1|U_2)=I(U_1;Y_1|U_2)+I(V_1;Y_1|U_1,U_2)    \\ \label{HK10}
	R_{U_2}+R_{V_2}   & ~~{\le}~~   I(U_2,V_2;Y_2|U_1)=I(U_2;Y_2|U_1)+I(V_2;Y_2|U_1,U_2)   \\ \label{HK11}
	R_{U_2}+R_{V_1}   & ~~{\le}~~  I(U_2,V_1;Y_1|U_1)=I(U_2;Y_1|U_1)+I(V_1;Y_1|U_1,U_2)
	   \\ \label{HK12}  
	R_{U_1}+R_{V_2}    & ~~{\le}~~   I(U_1,V_2;Y_2|U_2)=
	I(U_1;Y_2|U_2)+I(V_2;Y_2|U_1,U_2)   \\  \label{HK13}
	R_{U_1}+R_{U_2}+ R_{V_1}  & ~~{\le}~~   I(U_1,U_2,V_1;Y_1)=
	 I(U_1,U_2;Y_1)+ I(V_1;Y_1|U_1,U_2)    \\ \label{HK14}
	R_{U_1}+R_{U_2} + R_{V_2}    & ~~{\le}~~   I(U_1,U_2,V_2;Y_2)=
	 I(U_1,U_2;Y_2)+ I(V_2;Y_2|U_1,U_2)   \\ \label{HK15}
	E(X_1^2)& ~~=~~   P_1  \\  \label{HK16}
	E(X_2^2)& ~~=~~   P_2.
\end{align}
Let us consider \ref{HK1p} to \ref{HK14} in conjunction with independent and identically distributed (single-letter) Gaussian code-books for $U_1$, $U_2$, $V_1$, $V_2$.  For Gaussian code-books, the degradedness properties established in Theorem~\ref{V4Th2} 	are valid. 
Since the above formulation results in an achievable weighted sum-rate, any set of restrictive assumptions, if consistent with \ref{HK1p} to \ref{HK16}, results in an achievable (potentially inferior) solution.    Let us restrict $U_1$, $U_2$, $V_1$, $V_2$  to  be independent, 
$X_1=U_1+V_1$, $X_2=U_2+V_2$. We have $E(X_1^2)=E(U_1^2)+ E(V_1^2)$ and $E(X_2^2)=E(U_2^2)+ E(V_2^2)$.
 For given power allocation and encoding/decoding strategies (determining the values of mutual information terms on right hand sides of \ref{HK1} to \ref{HK14}), optimization problem in \ref{HK1p} to \ref{HK14} will be a linear programming problem with four variables, i.e.,  $R_{U_1}$, $R_{U_2}$, $R_{V_1}$, $R_{V_2}$.   This means, in the optimum solution, at least 4 constraints among  \ref{HK1} to \ref{HK14} will be satisfied with equality, resulting in zero value for  the corresponding slack variables\footnote{It turns out, with optimized  power allocation and encoding/decoding strategies, a higher number of slack variables may become zero. In view of the dual linear program, these additional zero-valued slack variables will be advantageous in increasing the value of the objective function.} 
 
 Let us further restrict the HK region by imposing  
\begin{equation}
	R_{V_1}  =  I(V_1;Y_1|U_1,U_2),~~~
	R_{V_2} =  I(V_2;Y_2|U_1,U_2),~~~
	R_{U_1}+R_{U_2}  =  I(U_1,U_2;Y_1) =  I(U_1,U_2;Y_2).
\end{equation} 
For such a restricted region, we have
\begin{align} 
	\label{sHK1p}
	\mbox{Maximize:}~ & R_1+\mu R_2=R_{U_1}+R_{V_1}+\mu R_{U_2}+\mu R_{V_2}~\mbox{where}\\ \label{sHK1}
	R_{U_1}  & ~~{\le}~~  I(U_1;Y_1|U_2) ~\stackrel{\text{(a)}}{\leq}~  I(U_1;Y_1|U_2,V_1)  \\ \label{sHK10}
	R_{U_1}    & ~~{\le}~~   I(U_1;Y_2|U_2)  ~\stackrel{\text{(b)}}{\leq}~ I(U_1;Y_2|U_2,V_2)  \\  \label{sHK13}
	R_{U_2}  & ~~{\le}~~  I(U_2;Y_1|U_1)  ~\stackrel{\text{(c)}}{\leq}~  I(U_2;Y_1|U_1,V_1) \\ \label{sHK12}  
	R_{U_2}  & ~~{\le}~~   I(U_2;Y_2|U_1) ~\stackrel{\text{(d)}}{\leq}~  I(U_2;Y_2|U_1,V_2)  \\ \label{sHK11}
	R_{U_1}+R_{U_2} & ~~\stackrel{\text{(e)}}{=}~I(U_1,U_2;Y_1)    \\ \label{sHK14}
	R_{U_1}+R_{U_2}   & ~~\stackrel{\text{(f)}}{=}~~   I(U_1,U_2;Y_2)    \\ \label{sHK15}
	R_{V_1}  & ~~{=}~~  I(V_1;Y_1|U_1,U_2)   \\ \label{sHK6}
	R_{V_2}  & ~~{=}~~   I(V_2;Y_2|U_1,U_2)    \\ \label{sHK7}  
	R_{U_1}    & ~~\stackrel{\text{(a)}}{\leq}~~   I(U_1;Y_1|U_2,V_1)     \\ \label{sHK2}
	R_{U_1}  & ~~\stackrel{\text{(b)}}{\leq}~~   I(U_1;Y_2|U_2,V_2)    \\ \label{sHK3}
	R_{U_2}   & ~~\stackrel{\text{(c)}}{\leq}~~    I(U_2;Y_1|U_1,V_1)    \\ \label{sHK4}
	R_{U_2}   & ~~\stackrel{\text{(d)}}{\leq}~~   I(U_2;Y_2|U_1,V_2)    \\  \label{sHK5}
	R_{U_1}+R_{U_2}   & ~~{\le}~~ I(U_1,U_2;Y_1|V_1) ~\stackrel{\text{(e)}}{\leq}~I(U_1,U_2;Y_1)   \\  \label{sHK8}
	R_{U_1}+R_{U_2}   & ~~{\le}~~   I(U_1,U_2;Y_2|V_2) ~\stackrel{\text{(f)}}{\leq}~I(U_1,U_2;Y_2) \\  \label{sHK9}
	E(X_1^2)& ~~=~~   P_1  \\  \label{sHK16}
	E(X_2^2)& ~~=~~   P_2.
\end{align}
Noting relationships specified by (a),(b),(c),(d),(e) and (f) in \ref{sHK1p} to \ref{sHK8}, it follows that \ref{sHK7} to  \ref{sHK8} are redundant. 
Upon removing redundant constraints from \ref{sHK1p} to  \ref{sHK16}, we obtain
\begin{align} 
	\label{zHK1ppz}
	\mbox{Maximize:}~~~~~ & R_1+\mu R_2 ~~~\mbox{where} \\ \label{zHK1}
	R_{U_1}  & ~~{\le}~~  I(U_1;Y_1|U_2)    \\ \label{zHK4}
	R_{U_1}    & ~~{\le}~~   I(U_1;Y_2|U_2)    \\  \label{zHK7}
	R_{U_2}  & ~~{\le}~~  I(U_2;Y_1|U_1)  \\ \label{zHK6ppp}  
	R_{U_2}  & ~~{\le}~~   I(U_2;Y_2|U_1)   \\ \label{zHK5}
	R_{U_1}+R_{U_2} & ~~{=}~~   I(U_1,U_2;Y_1)    \\ \label{zHK8}
	R_{U_1}+R_{U_2}   & ~~{=}~~   I(U_1,U_2;Y_2)    \\ \label{zHK9}
		R_{V_1}  & ~~{=}~~  I(V_1;Y_1|U_1,U_2)   \\ \label{zHK2}
	R_{V_2}  & ~~{=}~~   I(V_2;Y_2|U_1,U_2)    \\ \label{zHK3}  
	E(X_1^2)& ~~=~~   P_1  \\  	\label{zHK10}
E(X_2^2)  & ~~=~~  P_2.
\end{align}
Let us consider the following problem: 
\begin{align} 
	\label{zHK1pp}
	\mbox{Maximize:}~~~~~ & R_1+\mu R_2 ~~~\mbox{where} \\ \label{zHK1p}
	R_{U_1}  & ~~{=}~~  I(U_1;Y_2|U_2)\stackrel{\text{(g)}}{\leq}  I(U_1;Y_1|U_2)   \\ \label{zHK4p}
	%	R_{U_1}    & ~~{\leq}~~  I(U_1;Y_2|U_2)     \\  \label{zHK7nn}
		R_{U_2}  & ~~{=}~~  I(U_2;Y_1)\leq I(U_2;Y_1|U_1) \stackrel{\text{(h)}}{\leq}~  I(U_2;Y_2|U_1) \\ \label{zHK66}  
		%	R_{U_2}  &  ~\leq~  I(U_2;Y_2|U_1)   \\ \label{zHK5p}
	R_{U_1}+R_{U_2} & ~~{=}~~    I(U_1,U_2;Y_1)    \\ \label{zHK88p}
	R_{U_1}+R_{U_2}   & ~~{=}~~   I(U_1,U_2;Y_2)    \\ \label{zHK9p}
	R_{V_1}  & ~~{=}~~  I(V_1;Y_1|U_1,U_2)   \\ \label{zHK2p}
	R_{V_2}  & ~~{=}~~   I(V_2;Y_2|U_1,U_2)    \\ \label{zHK3p}  
	E(X_1^2)& ~~=~~   P_1  \\  	\label{zHK10p}
	E(X_2^2)  & ~~=~~  P_2
\end{align}
where $(g)$ in \ref{zHK1pp} and $(h)$ in \ref{zHK4p} are from Theorem \ref{V4Th2}. 
The solution to  \ref{zHK1pp} to \ref{zHK10p} is potentially inferior to the solution of the problem in \ref{zHK1ppz} to \ref{zHK10}, since the feasible region is shrunk by enforcing equality/inequality conditions in \ref{zHK1p} and \ref{zHK4p}.
Simplifying  \ref{zHK1pp} to \ref{zHK10p} results in
\begin{align} 
	\label{zHK1ppZZAA}
	\mbox{Maximize:}~~~~~ & R_1+\mu R_2 ~~~\mbox{where} \\ \label{zHK1pZZAA}
	R_{U_1}  & ~~{=}~~  I(U_1;Y_1|U_2)  \\ \label{zHK4pZZAA}
	R_{U_2}  & ~~{=}~~   I(U_2;Y_1)   \\ \label{zHK5pZZAA}
	R_{U_1}+R_{U_2}   & ~~{=}~~   I(U_1,U_2;Y_2) ~=~I(U_1,U_2;Y_1)   \\ \label{zHK9pZZAA}
	R_{V_1}  & ~~{=}~~  I(V_1;Y_1|U_1,U_2)   \\ \label{zHK2pZZAA}
	R_{V_2}  & ~~{=}~~   I(V_2;Y_2|U_1,U_2)    \\ \label{zHK3pZZAA}  
	E(X_1^2)& ~~=~~   P_1  \\  	\label{zHK10pZZAA}
	E(X_2^2)  & ~~=~~  P_2.
\end{align}
 Solution to \ref{zHK1ppZZAA} to \ref{zHK10pZZAA} results: (1) an achievable region potentially inferior to the HK region due to shrinking of the corresponding  feasible region, and (2) the solution  coincides with optimum boundary established in Section~\ref{S5} for $\mu<1$. Similar arguments can be applied to $\mu>1$. This entails Han-Kobayashi region with independent and identically distributed (single-letter) Gaussian code-books for $U_1$, $U_2$, $V_1$, $V_2$  is optimum.

\begin{center}
	{\bf \LARGE Appendix}
\end{center}
\begin{appendix}

In the following, to simplify expressions, entropy values are computed in base ``e". 
%
%$$
%H_1( \mathtt{u}_1+\mathtt{v}_1
%	+\sqrt{b}\mathtt{u}_2+\sqrt{b}\mathtt{v}_2+\mathtt{z}_1)
%-
%H_4(\mathtt{v}_1+
%	\sqrt{b}\mathtt{v}_2+\mathtt{z}_1)=H_7( \sqrt{a}\mathtt{u}_1+\sqrt{a}\mathtt{v}_1
%	+\mathtt{u}_2+\mathtt{v}_2+\mathtt{z}_2)
%	-
%H_8(\sqrt{a}\mathtt{v}_1+\mathtt{v}_2+\mathtt{z}_2)
%$$
\section{Proof of Theorem \ref{S4NT}:  General Formulation for Message Code-books} \label{App0}

{\bf Claim:} {\em      Fixing $\|\mathtt x_1\|^2$, $\|\mathtt x_2\|^2$, any (local/global) optimum solution  maximizing $\mathbi{R}_1 +\mu 	\mathbi{R}_2$ should rely on Gaussian densities for  $\mathtt x_1$, $\mathtt x_2$.}

\noindent 
{\bf Proof:} Provided Appendices \ref{App1} and \ref{LIC2}.  ~~~~$\square$

\subsection{Optimality of Gaussian  Code-books for General Vector Messages} \label{App1}
Consider vector inputs $\mathbf x_1$, $\mathbf x_2$ resulting in $\mathbf y_1=\mathbf x_1+\sqrt{b} \mathbf x_2+\mathbf z_1$ and $\mathbf y_2= \mathbf x_2+\sqrt{a}\mathbf x_1+\mathbf z_2$.  Since $\mathbf x_1$ and $\mathbf x_2$ are independent, we have 
\begin{eqnarray} \label{Eq201}
	\mathbi{R}_1  & = & H_{\mathtt{y}_1}(\mathbf x_1+\sqrt{b} \mathbf x_2+\mathbf z_1)-H_{\mathtt{n}_1}( \sqrt{b} \mathbf x_2+\mathbf z_1) \\ \label{Eq202}
	\mathbi{R}_2  & = & H_{\mathtt{y}_2}( \mathbf x_2+\sqrt{a}\mathbf x_1+\mathbf z_2)-H_{\mathtt{n}_2}( \sqrt{a} \mathbf x_1+\mathbf z_2).
\end{eqnarray}
Given power values $\|\mathbf x_1\|^2$  and $\|\mathbf x_2\|^2$, consider the LICQ conditions (in terms of perturbing densities of $\mathbf x_1$ and $\mathbf x_2$).  LICQ is satisfied if the determinant of matrix 
\begin{equation}
	\begin{bmatrix}
		\Big| \delta H_{\mathtt{y}_1}[\Delta \mathbf x_1] &  
		\delta H_{\mathtt{y}_1}[\sqrt{b},\Delta \mathbf x_2]-\delta H_{\mathtt{n}_1}[ \sqrt{b},\Delta \mathbf x_2]\\
		\delta H_{\mathtt{y}_2}[\sqrt{a},\Delta \mathbf x_1]-\delta H_{\mathtt{n}_2}[\sqrt{a},\Delta \mathbf x_1]   & \delta H_{\mathtt{y}_2}[\Delta \mathbf x_2] \Big|
	\end{bmatrix} \label{Eq203}
\end{equation}
Determinant of the matrix in \ref{Eq203} will be non-zero if 
\begin{eqnarray} \label{Eq204}
	\frac{\delta H_{\mathtt{y}_2}[\sqrt{a},\Delta \mathbf x_1]-\delta H_{\mathtt{n}_2}[\sqrt{a},\Delta \mathbf x_1]}
	{\delta \delta H_{\mathtt{y}_1}[\Delta \mathbf x_1]}
	\neq 
	\frac{\delta H_{\mathtt{y}_2}[\Delta \mathbf x_2]}{\delta H_{\mathtt{y}_1}[\sqrt{b},\Delta \mathbf x_2]-\delta H_{\mathtt{n}_1}[ \sqrt{b},\Delta \mathbf x_2]}.
\end{eqnarray}
Subject to \ref{Eq204}, a necessary condition for any local/global optimum solution is that the first order variations of  $\mathbi{R}_1+\mu \mathbi{R}_2$ should be zero. This means
\begin{eqnarray} \label{Eq205p}
	\delta_{\mathbi{R}_1+\mu \mathbi{R}_2}[\Delta  \mathbf x_1] &=& \delta H_{\mathtt{y}_1}[\Delta \mathbf x_1]+\mu \Big({\delta H_{\mathtt{y}_2}[\sqrt{a},\Delta \mathbf x_1]-\delta H_{\mathtt{n}_2}[\sqrt{a},\Delta \mathbf x_1]}\Big) ~=~0\\  \label{Eq206p}
	\delta_{\mathbi{R}_1+\mu \mathbi{R}_2}[\Delta  \mathbf x_2] &=& 
	\Big(\delta H_{\mathtt{y}_1}[\sqrt{b},\Delta \mathbf x_2]-\delta H_{\mathtt{n}_1}[ \sqrt{b},\Delta \mathbf x_2]\Big)+\mu \delta H_{\mathtt{y}_2}[\Delta \mathbf x_2]~=~0.
\end{eqnarray}
From \ref{Eq205p} and  \ref{Eq206p}, we obtain
\begin{eqnarray} \label{Eq204z}
	\frac{\delta H_{\mathtt{y}_2}[\sqrt{a},\Delta \mathbf x_1]-\delta H_{\mathtt{n}_2}[\sqrt{a},\Delta \mathbf x_1]}
	{\delta \delta H_{\mathtt{y}_1}[\Delta \mathbf x_1]}=
	\frac{\delta H_{\mathtt{y}_2}[\Delta \mathbf x_2]}{\delta H_{\mathtt{y}_1}[\sqrt{b},\Delta \mathbf x_2]-\delta H_{\mathtt{n}_1}[ \sqrt{b},\Delta \mathbf x_2]}=\psi.
\end{eqnarray}
Since \ref{Eq204} contradicts \ref{Eq204z}, it follows that \ref{Eq204z} is valid at any local/global optimum solution. 
Combining \ref{Eq204z} with \ref{Eq205p} and \ref{Eq206p} results in 
\begin{eqnarray} \label{Eq207p}
	\delta_{\mathbi{R}_1+\mu \mathbi{R}_2}[\Delta  \mathbf x_1] &=& 
	(1+\mu \Psi)\,\delta H_{\mathtt{y}_1}[\Delta \mathbf x_1]  \\  \label{Eq208p}
	\delta_{\mathbi{R}_1+\mu \mathbi{R}_2}[\Delta  \mathbf x_2] &=& 
	(\Psi+\mu)\,\Big(\delta H_{\mathtt{y}_1}[\sqrt{b},\Delta \mathbf x_2]-\delta H_{\mathtt{n}_1}[ \sqrt{b},\Delta \mathbf x_2]\Big).
\end{eqnarray}
Note that \ref{Eq207p} and \ref{Eq208p} linearly scale with the step sizes 
$\epsilon_{{\mathbf x}_1}$ and $\epsilon_{{\mathbf x}_2}$, respectively. This means  
\begin{equation}
	\Big(\delta_{\mathbi{R}_1+\mu \mathbi{R}_2}[\Delta  \mathbf x_1], \delta_{\mathbi{R}_1+\mu \mathbi{R}_2}[\Delta  \mathbf x_2]\Big)
\end{equation} 
spans a two-dimensional space. Consequently, any perturbation of densities around a local/global optimum solution changes the value of $\mathbi{R}_1+\mu \mathbi{R}_2$.  To have a local/global optimum solution, we should have
\begin{eqnarray} \label{Eq207z}
	\delta H_{\mathtt{y}_1}[\Delta \mathbf x_1] & = & 0 \\  \label{Eq208z}
	\delta H_{\mathtt{y}_1}[\sqrt{b},\Delta \mathbf x_2]-\delta H_{\mathtt{n}_1}[ \sqrt{b},\Delta \mathbf x_2] & = & 0
\end{eqnarray}
subject to the constraints on ``area under densities'' and ``power values''.  From \ref{Eq207z}, it is concluded that  $\mathbf y_1=\mathbf x_1+\sqrt{b} \mathbf x_2+\mathbf z_1$ is Gaussian. Noting $\mathbf x_1$, $\mathbf x_2$,  $\mathbf z_1$ are independent,  by  applying Cramér’s theorem ~\cite{GN1}\cite{GN2}, it follows that $\mathtt{x}_1$ and $\mathtt{x}_2$ should be Gaussian.

\subsection{Gaussian is the Unique Density Satisfying the First-Order Optimality Condition}
\label{LIC2}
Without loss of generality, let us consider scalar versions of the  code-books. Generalization to vectors is straightforward if the constraint on power is  replaced by constraint on the  vector covariance  matrix. 
We show that among all probability densities \(f(x)\) on \(\mathbb{R}\) with fixed mean and fixed variance (or fixed covariance matrix in the vector case), the Gaussian density maximizes the differential entropy.
The differential entropy in scalar case is
\begin{equation}
	h(f) = -\int_{-\infty}^{\infty} f(x)\log f(x)\,dx.
	\label{E173}
\end{equation}
Assume the constraints
\begin{eqnarray} 	\label{E174}
	\int_{-\infty}^{\infty} f(x)\,dx & = & 1 \\ 	\label{E175}
	\int_{-\infty}^{\infty} x f(x)\,dx & = & m \\ 	\label{E176}
	\int_{-\infty}^{\infty} x^2 f(x)\,dx & = &\sigma^2+m^2.
\end{eqnarray}
Define the Lagrangian functional
\begin{equation}
	\mathcal{L}[f]
	\!=	\!-\!\int\!f(x)\log f(x)dx
	-\!\lambda_0\!\left(\int\!f(x)\,dx-\!1\right)\!-\!\ \lambda_1\!\left(\int\!\!x\!f(x)dx\!-\!m\right)
	\!-\!\lambda_2\!\left(\int\!x^2 f(x)dx\!-\!m^2 \!-\!\sigma^2\right).
\end{equation}
To find a stationary point, perturb \(f\) as
\begin{equation}
	f_\epsilon(x) = f(x) + \epsilon \eta(x)~~\mbox{where}~~\int\eta(x)dx=0.
\end{equation}
The first variation condition is
\begin{equation}
	\left.
	\frac{d}{d\epsilon}
	\mathcal{L}[f + \epsilon \eta]
	\right|_{\epsilon=0}.
	= 0
\end{equation}
\begin{equation}
	\frac{d}{d\epsilon}
	\Big(
	\!-\!(f+\epsilon\eta)\log(f+\epsilon\eta)
	\Big)
	=
	-\eta\,\Big(\log(f+\epsilon\eta)+1\Big).
\end{equation}
Therefore, at \(\epsilon = 0\), we have
\begin{equation}
	\delta \mathcal{L}[f;\eta]
	=
	\int
	\Big(
	\!\!-\!\log f(x)
	-1
	-\lambda_0
	-\lambda_1 x
	-\lambda_2 x^2
	\Big)
	\eta(x)\,dx.
	\label{N291}
\end{equation}
For \ref{N291} to vanish for all admissible perturbations \(\eta(x)\), we must have
\begin{equation}
	-\log f(x)
	-1
	-\lambda_0
	-\lambda_1 x
	-\lambda_2 x^2
	= 0.
\end{equation}
Noting $\eta(x)$ integrates to zero, we must have
\begin{equation}
	\log f(x)
	=
	-1-\lambda_0-\lambda_1 x
	-\lambda_2 x^2.
\end{equation}
Thus
\begin{equation}
	f(x)
	=
	C_2
	\exp
	\left[
	-\lambda_2
	\left(
	x + \frac{\lambda_1}{2\lambda_2}
	\right)^2
	\right].
\end{equation}
Matching the mean and variance gives
\begin{equation}
	f^\star(x)
	=
	\frac{1}{\sqrt{2\pi\sigma^2}}
	\exp
	\left[
	-\frac{(x-m)^2}{2\sigma^2}
	\right].
\end{equation}
This means Gaussian is a stationary (local optimum) solution. Next, we need to show that Gaussian is globally optimum. 
The second variation of entropy is
\begin{equation}
	\delta^2 h[f;\eta]
	=
	-\int
	\frac{\Big(\eta(x)\Big)^2}{f(x)}
	\,dx.
\end{equation}
Since \(f(x) > 0\), we have
\begin{equation}
	\delta^2 h[f;\eta] \le 0.
\end{equation}
with equality only when
\begin{equation}
	\eta(x) = 0.
\end{equation}
Therefore, the entropy functional is strictly concave in \(f\) and 
hence has a unique maximizing solution.  Consequently, given   mean \(m\) and variance \(\sigma^2\), the Gaussian density uniquely maximizes the entropy (see \cite{EPFL1} (page 5) and references therein for alternative proofs).  Finally, noting that it is necessary for the first-order variation to vanish, it is concluded that the Gaussian distribution is the unique density for which the first-order variation is zero. 
\section{Formation of Rate Expressions } \label{LIC1}
\subsection{Entropy Term Involving a Convolution of Density Functions} \label{AppA}
Consider 
\begin{equation}
	f_S(s)=f_V(v)\ast f_W(w)\ast f_Z(z)=
	\int_v\int_w f_V(v)f_W(w)f_Z(s-v-w)dw dv. 
	\label{E1}
\end{equation}
Perturbing $f_V(v)$ changes the 
entropy term to
\begin{eqnarray}
	\int_s \Big[
	\int_v \int_w \Big(f_V(v)+\epsilon_V \Lambda_V (v)\Big)f_W(w)f_Z(s-w-v) dw dv \times \\
	\ln\int_v \int_w \Big(f_V(v)+\epsilon_V \Lambda_V (v)\Big)f_W(w)f_Z(s-w-v) dw dv\Big]ds.
\end{eqnarray}
Corresponding variation is equal to
\begin{eqnarray} \label{E199}
	\epsilon_V \int_s \Big[\Big(1+	\ln \int_v\int_w f_V(v)f_W(w)f_Z(s-v-w)dw dv\Big)
	\int_v \int_w \Lambda_V (v)f_W(w)f_Z(s-w-v) dw dv \Big] ds =\\ \label{E200}
	\epsilon_V \int_s \Big[\Big(1+	\ln f_S(s)\Big)
	\int_v \int_w \Lambda_V (v)f_W(w)f_Z(s-w-v) dw dv \Big] ds.~~~ 
\end{eqnarray}
\section{Proof of Theorem \ref{V4Th5p}: HK Formulation with Pubic/Private Messages} \label{AppB}

{\bf Claim:} {\em      A necessary condition for optimality of a set of vector code-books is that the first-order variation of the weighted sum-rate vanishes; and this condition can be satisfied  only if  code-books are Gaussian.}

\noindent 
{\bf Proof:}
Consistent with arguments regrading range of $\mu<1$ in Theorem \ref{V4Th5z}, we have
\begin{eqnarray} 	\label{ApEq1ppnN}
	\mathbf{r}_{\mathtt{u}_2}
	& = &
	\overbrace{H_1( \mathtt{u}_1+\mathtt{v}_1
	+\sqrt{b}\mathtt{u}_2+\sqrt{b}\mathtt{v}_2+\mathtt{z}_1)}^{\mathtt{s}_{\mathtt{u}_2}}
	-
		\overbrace{H_3( \mathtt{u}_1
	+\mathtt{v}_1+
	\sqrt{b}\mathtt{v}_2+\mathtt{z}_1)}^{\mathtt{n}_{\mathtt{u}_2}\equiv\,\mathtt{s}_{\mathtt{u}_1}}
	\\ \label{ApEq2ppnN}
	\mathbf{r}_{\mathtt{u}_1}
	& = &
		\overbrace{H_3( \mathtt{u}_1+\mathtt{v}_1
	+\sqrt{b} \mathtt{v}_2+\mathtt{z}_1)}^{\mathtt{s}_{\mathtt{u}_1}}
	-
		\overbrace{H_4( \mathtt{v}_1
	+\sqrt{b} \mathtt{v}_2+\mathtt{z}_1)}^{\mathtt{n}_{\mathtt{u}_1}}
	\\  \label{ApEq3ppnN}
	\mathbf{r}_{\mathtt{v}_1}
	& = &
			\overbrace{H_4(\mathtt{v}_1+
	\sqrt{b}	\mathtt{v}_2+
	\mathtt{z}_1)}^{\mathtt{s}_{\mathtt{v}_1}\equiv\, \mathtt{n}_{\mathtt{u}_1}}
	-
				\overbrace{H_2( \sqrt{b}\mathtt{v}_2+
					\mathtt{z}_1)}^{\mathtt{n}_{\mathtt{v}_1}} \\  \label{ApEq4ppnN}
	\mathbf{r}_{\mathtt{v}_2}  & = & 	\overbrace{H_5(\sqrt{a}	\mathtt{v}_1+\mathtt{v}_2+
		\mathtt{z}_2)}^{\mathtt{s}_{\mathtt{v}_2}}
	-\overbrace{H_6( \sqrt{a}\mathtt{v}_1+\mathtt{z}_2)}^{\mathtt{n}_{\mathtt{v}_2}}
\end{eqnarray} 
where subscripts are added to entropy terms, i.e., $H_1$ to $H_6$, to distinguish different compound random vectors.   
%In dealing with  \ref{ApEq1ppn} to \ref{ApEq4ppn}, a compound random variable may appear in two active entropy terms. This occurs in successive decoding, where the noise  in a layer reappears as the signal  in another  layer. This scenario can be avoided  noting that expressions 
Expressions  in \ref{ApEq1ppnN} to \ref{ApEq4ppnN} can be changed to the following equivalent set
\begin{eqnarray} 	\label{Eq196}
	\mathbf{r}_{\mathtt{u}_2}+\mathbf{r}_{\mathtt{u}_1}+\mathbf{r}_{\mathtt{v}_1}
	& = &
	\overbrace{H_1( \mathtt{u}_1+\mathtt{v}_1
		+\sqrt{b}\mathtt{u}_2+\sqrt{b}\mathtt{v}_2+\mathtt{z}_1)}^{\mathtt{s}_{\mathtt{u}_2}}
	-
	\overbrace{H_2( \sqrt{b}\mathtt{v}_2+\mathtt{z}_1)}^{\mathtt{n}_{\mathtt{v}_1}}
	\\ \label{Eq197}
	\mathbf{r}_{\mathtt{u}_1}+	\mathbf{r}_{\mathtt{v}_1}
	& = &
	\overbrace{H_3( \mathtt{u}_1+\mathtt{v}_1
		+\sqrt{b} \mathtt{v}_2+\mathtt{z}_1)}^{\mathtt{s}_{\mathtt{u}_1}}
	-
	\overbrace{H_2( \sqrt{b}\mathtt{v}_2+\mathtt{z}_1)}^{\mathtt{n}_{\mathtt{v}_1}}
	\\  \label{Eq198}
	\mathbf{r}_{\mathtt{v}_1}
	& = &
	\overbrace{H_4(\mathtt{v}_1+
		\sqrt{b}	\mathtt{v}_2+
		\mathtt{z}_1)}^{\mathtt{s}_{\mathtt{v}_1}}
	-
	\overbrace{H_2( \sqrt{b}\mathtt{v}_2+\mathtt{z}_1)}^{\mathtt{n}_{\mathtt{v}_1}} \\ \label{Eq199}
		\mathbf{r}_{\mathtt{v}_2}
	& = &
	\overbrace{	H_5(	\sqrt{a} \mathtt{v}_1+\mathtt{v}_2+
		\mathtt{z}_2)}^{\mathtt{s}_{\mathtt{v}_2}}
	-
	\overbrace{H_6( \sqrt{a}\mathtt{v}_1+\mathtt{z}_2)}^{\mathtt{n}_{\mathtt{v}_2}}
\end{eqnarray}
Without loss of generality, let us express LICQ in terms of scalar code-books. The arguments presented here can also be interpreted in terms of the Jacobian matrix. The validity of generalization to vectors will be explained in relation to some key steps in establishing  the conditions for scalar code-books. Readers are referred to Appendices \ref{LIC2} and \ref{LIC1} for  the effect of constraints on power and area under densities. 
Let us include the scaling factors $\sqrt{a}$ and $\sqrt{b}$ as additional arguments of the entropy functional. Vectors to be linearly combined in verifying LICQ in conjunction with the scalar version of \ref{Eq196} to \ref{Eq199}  are as follows:
\begin{eqnarray}  \label{Eq202z}
\begin{bmatrix}
	\overbrace{\delta H_1[\Delta U_1]}^{I_1}  & 	\!\overbrace{\delta H_1[\Delta V_1]}^{I_2}  & 	\!\!\!\!\!\!\!\!\overbrace{\delta H_1[\sqrt{b},\Delta U_2]}^{I_3}~~~~  & 	
		\overbrace{\delta H_1[\sqrt{b},\Delta V_2]\!-\!\delta H_2[\sqrt{b},\Delta V_2]}^{I_4} \\
			\overbrace{\delta H_3[\Delta U_1]}^{I_5}  & 		\overbrace{\delta H_3[\Delta V_1]}^{I_6}  & 	0 & 	\overbrace{\delta H_3[\sqrt{b},\Delta V_2]\!-\!\delta H_2[\sqrt{b},\Delta V_2]}^{I_7} \\
						0 & 	\overbrace{\delta H_4[\Delta V_1]}^{I_8}  & 	0 &  		\overbrace{\delta H_4[\sqrt{b},\Delta V_2]\!-\!\delta H_2[\sqrt{b},\Delta V_2]}^{I_9}   \\
						0 &  \overbrace{\delta H_5[\sqrt{a},\Delta V_1]\!-\!\delta H_6[\sqrt{a},\Delta V_1]}^{I_{10}} &  0 &  \overbrace{\delta H_5[\Delta V_2]}^{I_{11}}
	\end{bmatrix}
	\begin{bmatrix}
			\alpha_{U_1} \\
				\alpha_{V_1} \\
							\alpha_{U_2}\\
				\alpha_{V_2}\\
			\end{bmatrix}~~~~~~~~~
\end{eqnarray}
\begin{eqnarray} \label{Eq202p}
	\left|\begin{bmatrix}
		I_1 & I_2 & I_3 & I_4 \\
		I_5 & I_6 & 0 & I_7 \\
		0 & I_8 & 0 & I_9 \\
		0 & I_{10}& 0 & I_{11} 
	\end{bmatrix}\right|=I_3 I_5 I_8 I_{11}
	-I_3 I_5 I_9 I_{10}.
\end{eqnarray}
Note that matrix in \ref{Eq202z} is the Jacobian for expressions on the right hand sides of \ref{Eq196} to \ref{Eq199}.  
To show that LICQ is satisfied, one needs to show that the matrix in \ref{Eq202z}/\ref{Eq202p} is not singular, i.e., 
\begin{eqnarray} \nonumber 
		\overbrace{\overbrace{\delta H_1[\sqrt{b},\Delta U_2]}^{I_3}\overbrace{\delta H_3[\Delta U_1]}^{I_5}}^{\Phi_1} \times          \\                                               
	\label{Eq203p}
		\!\Big[\overbrace{\overbrace{\delta H_4[\Delta V_1]}^{I_8}\,\overbrace{\delta H_5[\Delta V_2]}^{I_{11}}}^{\Phi_2}-\overbrace{\Big(\overbrace{\delta H_4[\sqrt{b},\Delta V_2]-\delta H_2[\sqrt{b},\Delta V_2]}^{I_9} \Big)
\Big(\overbrace{\delta H_5[\sqrt{a},\Delta V_1]-\delta H_6[\sqrt{a},\Delta V_1]}^{I_{10}}}^{\Phi_3}\Big)\Big]\neq 0.
\end{eqnarray}
 In \ref{Eq202p}/\ref{Eq203p}, we have $\delta H_1[\sqrt{b},\Delta U_2]\neq 0$ and  
 $\delta H_3[\Delta U_1]\neq 0$, resulting in $\Phi_1\neq 0$. 
 Justification for $\delta H_1[\sqrt{b},\Delta U_2]\neq 0$, as an example, is provided in \ref{Eq205} (also see Appendix \ref{LIC1}). 
\begin{eqnarray} \label{Eq205}
\delta H_1[\sqrt{b},\Delta U_2]  = \\
\frac{\epsilon_{U_2}}{b}\! \int (1+\log s)\Big[\!\int \int \int \int \! f_{U_1}f_{V_1}\Lambda_{\frac{U_2}{\sqrt{b}}}f_{\frac{V_2}{\sqrt{b}}}f_Z\Big(s-u_1-v_1-\frac{u_2}{\sqrt{b}}-\frac{v_2}{\sqrt{b}}\Big)du_1
dv_1du_2dv_2\Big]ds\neq 0.
\end{eqnarray}
This property, i.e., $\Phi_1\neq 0$, is valid for vector code-books as well. 
%Adding \ref{ApEq2ppnN} to \ref{ApEq3ppnN} and \ref{ApEq1ppnN} to \ref{ApEq4ppnN}, we obtain
%\begin{eqnarray} \label{Eq166}
%	I_8  =  \delta R_{V_1}[\Delta V_1],~~
%		I_9  = \delta R_{V_1}[\Delta V_2],~~
%			I_{10}  =  \delta R_{V_2}[\Delta V_1],~~
%		I_{11}  =   \delta R_{V_2}[\Delta V_2].
%\end{eqnarray}
Since, in \ref{Eq203p},   $\Phi_1\neq 0$, the condition is violated if 
\begin{eqnarray} \label{Eq206}
\overbrace{\delta H_4[\Delta V_1]\delta H_5[\Delta V_2]}^{\Phi_2})
& = &
\overbrace{
	\Big(\delta H_4[\sqrt{b},\Delta V_2]-\delta H_2[\sqrt{b},\Delta V_2]\Big)
	\Big(\delta H_5[\sqrt{a},\Delta V_1]-\delta H_6[\sqrt{a},\Delta V_1]\Big)}^{\Phi_3} \\  \label{Eq207}
	\frac{I_8}{I_{10}} & = & \frac{I_9}{I_{11}}	~~\Longrightarrow~~
	\frac{\delta R_{V_1}[\Delta V_1]}{\delta R_{V_2}[\Delta V_1]}=	
	\frac{\delta R_{V_1}[\Delta V_2]}{\delta R_{V_2}[\Delta V_2]}\equiv \zeta,~~\forall \epsilon_{V_1}\simeq 0,~\forall \epsilon_{V_2}\simeq 0. 
\end{eqnarray}
Note that $\epsilon_{V_1}$ and $\epsilon_{V_2}$ appear as scale factors and cancel from numerators and     denominators in \ref{Eq207}, hence the inclusion of  $\forall \epsilon_{V_1}\simeq 0$, $\forall \epsilon_{V_2}\simeq 0$.  This property is valid for vector code-books as well. 
The condition is violated, 
{\em if  there exist acceptable$^4$ perturbations $\Delta V_1\neq 0$ and $\Delta V_2\neq 0$, satisfying \ref{Eq207}, which do not change  the value of $R_1+\mu R_2$}. To answer this question, consider the space of all valid density functions for public and private messages, in conjunction with all acceptable perturbations\footnote{A perturbation $\Lambda(x)$ is acceptable if 
	$\displaystyle \int_x\!\Lambda(x)dx=0$ and  $\displaystyle\!\int_x \!x^2\Lambda(x)dx=0$.} of $\Delta V_1$ and $\Delta V_2$. Consider a point corresponding to (arbitrary) density functions $f_{U_1}$, $f_{V_1}$, $f_{U_2}$, $f_{V_2}$ with fixed second moments $P_{U_1}$, $P_{V_1}$, $P_{U_2}$, $P_{V_2}$ and (arbitrary) acceptable$^4$ perturbations $\Delta V_1$, $\Delta V_2$ satisfying \ref{Eq207}.  
We have 
\begin{eqnarray} \label{EE170}
	\delta_{R_1+\mu R_2}[\Delta V_1] & = &	 \delta R_{V_2}[\Delta V_1]\overbrace{\Big(\frac{\delta R_{U_1}[\Delta V_1]}{\delta R_{V_2}[\Delta V_1]}+\psi+\mu \frac{\delta R_{U_2}[\Delta V_1]}{\delta R_{V_2}[\Delta V_1]}+\mu\Big)}^{A} \\  \label{EE171}
	\delta_{R_1+\mu R_2}[\Delta V_2] & = &	 \delta R_{V_2}[\Delta V_2]\overbrace{\Big(\frac{\delta R_{U_1}[\Delta V_2]}{\delta R_{V_2}[\Delta V_2]}+\psi+\mu \frac{\delta R_{U_2}[\Delta V_2]}{\delta R_{V_2}[\Delta V_2]}+\mu\Big)}^{B}~~~\mbox{where} 
\end{eqnarray}
\vspace{-0.5cm}
\begin{eqnarray} \nonumber
	\delta R_{U_1}[\Delta V_1] & = &
	\frac{\epsilon_{V_1}}{b}\!\!\int (1+\log s)\Big[\!\int \!\!\int \!\!\int\!\! f_{U_1}\Lambda_{V_1}f_{\frac{V_2}{\sqrt{b}}}f_Z\Big(s-u_1-v_1-\frac{v_2}{\sqrt{b}}\Big)du_1
	dv_1dv_2\Big]ds\\ \label{E172}
	& - &
	\frac{\epsilon_{V_1}}{b}\!\!\int (1+\log s)\Big[\!\int \!\!\int \!\! \Lambda_{V_1}f_{\frac{V_2}{\sqrt{b}}}f_Z\Big(s-v_1-\frac{v_2}{\sqrt{b}}\Big)
	dv_1dv_2\Big]ds\\ \nonumber
	\delta R_{U_1}[\Delta V_2] & = &
	\frac{\epsilon_{V_2}}{b}\!\!\int (1+\log s)\Big[\!\int \!\!\int \!\!\int\!\! f_{U_1}f_{V_1}\Lambda_{\!\frac{V_2}{\sqrt{b}}}f_Z\Big(s-u_1-v_1-\frac{v_2}{\sqrt{b}}\Big)du_1
	dv_1dv_2\Big]ds\\  \label{E173}
	& - &
	\frac{\epsilon_{V_2}}{b}\!\!\int (1+\log s)\Big[\!\int \!\!\int \!\! f_{V_1}\Lambda_{\!\frac{V_2}{\sqrt{b}}}f_Z\Big(s-v_1-\frac{v_2}{\sqrt{b}}\Big)
	dv_1dv_2\Big]ds\\ \nonumber
	\delta R_{U_2}[\Delta V_1] & = &
	\frac{\epsilon_{V_1}}{b}\!\!\int (1+\log s)\Big[\!\int \!\!\int \!\!\int\!\!\int\!\! f_{U_1}\Lambda_{V_1}f_{\!\frac{U_2}{\sqrt{b}}}f_{\frac{V_2}{\sqrt{b}}}f_Z\Big(s-u_1-v_1-\frac{u_2}{\sqrt{b}}-\frac{v_2}{\sqrt{b}}\Big)du_1
	dv_1du_2dv_2\Big]ds \\  \label{E174}
	& - & 	\frac{\epsilon_{V_1}}{b}\!\!\int (1+\log s)\Big[\!\int \!\!\int \!\!\int\!\! f_{U_1}\Lambda_{V_1}f_{\!\frac{V_2}{\sqrt{b}}}f_Z\Big(s-u_1-v_1-\frac{v_2}{\sqrt{b}}\Big)du_1
	dv_1dv_2\Big]ds \\ \nonumber 
	\delta R_{U_2}[\Delta V_2] & = &
	\frac{\epsilon_{V_2}}{b}\!\!\int (1+\log s)\Big[\!\int \!\!\int \!\!\int\!\!\int\!\! f_{U_1}f_{V_1}f_{\!\frac{U_2}{\sqrt{b}}}\Lambda_{\!\frac{V_2}{\sqrt{b}}}f_Z\Big(s-u_1-v_1-\frac{u_2}{\sqrt{b}}-\frac{v_2}{\sqrt{b}}\Big)du_1
	dv_1du_2dv_2\Big]ds \\  \label{E175}
	& - & 	\frac{\epsilon_{V_2}}{b}\!\!\int (1+\log s)\Big[\!\int \!\!\int \!\!\int\!\! f_{U_1}f_{V_1}\Lambda_{\!\frac{V_2}{\sqrt{b}}}f_Z\Big(s-u_1-v_1-\frac{v_2}{\sqrt{b}}\Big)du_1
	dv_1dv_2\Big]ds \\ \nonumber
	\delta R_{V_1}[\Delta V_1] & = &
	\frac{\epsilon_{V_1}}{b}\!\!\int (1+\log s)\Big[\!\int \!\!\int \!\!\int\!\! \Lambda_{V_1}f_{\frac{V_2}{\sqrt{b}}}f_Z\Big(s-v_1-\frac{v_2}{\sqrt{b}}\Big)		dv_1dv_2\Big]ds\\ \label{E180}
	& - &
	\frac{\epsilon_{V_1}}{b}\!\!\int (1+\log s)\Big[\!\int \!\! f_{\!\frac{V_2}{\sqrt{b}}}f_Z\Big(s-\frac{v_2}{\sqrt{b}}\Big) dv_2\Big]ds\\ \nonumber
	\delta R_{V_2}[\Delta V_1] & = &
	\frac{\epsilon_{V_1}}{a}\!\!\int (1+\log s)\Big[\!\int \!\!\int \!\!\int\!\! \Lambda_{\!\frac{V_1}{\sqrt{a}}}f_{V_2}f_Z\Big(s-\frac{v_1}{\sqrt{a}}-v_2\Big)		dv_1dv_2\Big]ds\\   \label{E176}
	& - &
	\frac{\epsilon_{V_2}}{b}\!\!\int (1+\log s)\left[\int \!\! f_{V_2}f_Z\Big(s-v_2\Big)dv_2\right] ds \\ \nonumber
	\delta R_{V_1}[\Delta V_2] & = &
	\frac{\epsilon_{V_2}}{b}\!\!\int (1+\log s)\Big[\!\int \!\!\int \!\!\int\!\! f_{V_1}\Lambda_{\frac{V_2}{\sqrt{b}}}f_Z\Big(s-v_1-\frac{v_2}{\sqrt{b}}\Big)		dv_1dv_2\Big]ds\\ \label{E182}
	& - &
	\frac{\epsilon_{V_2}}{b}\!\!\int (1+\log s)\Big[\!\int \!\! f_{\!\frac{V_2}{\sqrt{b}}}f_Z\Big(s-\frac{v_2}{\sqrt{b}}\Big) dv_2\Big]ds\\ 
	\nonumber
	\delta R_{V_2}[\Delta V_2] & = &
	\frac{\epsilon_{V_2}}{a}\!\!\int (1+\log s)\Big[\!\int \!\!\int \!\!\int\!\! f_{\!\frac{V_1}{\sqrt{a}}}\Lambda_{V_2}f_Z\Big(s-\frac{v_1}{\sqrt{a}}-v_2\Big)		dv_1dv_2\Big]ds\\   \label{E177}
	& - &
	\frac{\epsilon_{V_2}}{a}\!\!\int (1+\log s)\left[\int \!\! f_{\!\frac{V_1}{\sqrt{a}}}f_Z\Big(s-v_2\Big)dv_2\right] ds 
\end{eqnarray}
Term $A$ in \ref{EE170} does not change with $\epsilon_{V_1}$, likewise term $B$ in \ref{EE171} does not change with $\epsilon_{V_2}$. However, term $\delta R_{V_2}[\Delta V_1]$ in  \ref{EE170} scales (linearly) with $\epsilon_{V_1}$, and term $\delta R_{V_2}[\Delta V_2]$ in \ref{EE171} scales (linearly) with 
$\epsilon_{V_2}$. This means  
$\delta_{R_1+\mu R_2}[\Delta V_1]\neq 0,\,\forall \epsilon_{V_1}\neq 0$ and
$\delta_{R_1+\mu R_2}[\Delta V_2]\neq 0,\,\forall \epsilon_{V_2}\neq 0$. This in turn entails, any perturbation $\Delta V_1\neq 0$ and/or $\Delta V_2\neq 0$, satisfying \ref{Eq207}, result in changing  the value of $R_1+\mu R_2$. The final conclusion is that moving around any feasible solution, including those  that are locally optimum, changes the value of $R_1+\mu R_2$. 

Next, we show that the vanishing of the first-order variation of the objective function,  i.e., 
\begin{eqnarray} \label{Eq174}
	\mathbf{r}_{\mathtt{u}_1}+	\mathbf{r}_{\mathtt{v}_1}+
	\mu\,	\mathbf{r}_{\mathtt{u}_2}+	\mu\, 	\mathbf{r}_{\mathtt{v}_2}
\end{eqnarray} 
implies that the random code-books should be Gaussian. This property follows from the special structure of \ref{ApEq1ppnN}–\ref{ApEq4ppnN}, in which $\mathtt{u}_2$ appears only in $H_1$ in \ref{ApEq1ppnN}, and $\mathtt{u}_1$ appears only in $H_3$ in \ref{ApEq2ppnN}. More formally, the expressions in \ref{ApEq1ppnN}–\ref{ApEq4ppnN} are linearly related to those in \ref{Eq196}–\ref{Eq199} as follows:
\begin{equation}
	\begin{bmatrix}
		\mathbf{r}_{\mathtt{u}_2}\\
		\mathbf{r}_{\mathtt{u}_1}\\
		\mathbf{r}_{\mathtt{v}_1}\\
		\mathbf{r}_{\mathtt{v}_2}
	\end{bmatrix}
	=
		\overbrace{
	\begin{bmatrix}
		1 & -1 & 0 & 0\\
		0 & 1 & -1 & 0\\
		0 & 0 & 1 & 0\\
		0 & 0 & 0 & 1
	\end{bmatrix}}^{\mathtt{T}}
	\begin{bmatrix}
		\mathbf{r}_{\mathtt{u}_2}+\mathbf{r}_{\mathtt{u}_1}+\mathbf{r}_{\mathtt{v}_1}\\
		\mathbf{r}_{\mathtt{u}_1}+\mathbf{r}_{\mathtt{v}_1}\\
		\mathbf{r}_{\mathtt{v}_1}\\
		\mathbf{r}_{\mathtt{v}_2}
	\end{bmatrix}.
\end{equation}
First order variations of \ref{Eq174}, with respect to  $\mathtt{u}_2$, $\mathtt{u}_1$, $\mathtt{v}_1$, $\mathtt{v}_2$,  is captured by components of the vector
 \begin{eqnarray} \label{Eq176}
 	\mathtt{T} \overbrace{
 	\begin{bmatrix}
 		I_1 & I_5 & 0 & 0 \\
 		I_2 & I_6 & I_8 & I_{10} \\
 		I_3 & 0 & 0 & 0 \\
 		I_4 & I_7 & I_9 & I_{11}
 	\end{bmatrix}}^{\mathtt{M}}
 	\begin{bmatrix}
 		\mu \\
 		1 \\
 		1 \\
 		\mu 
 	\end{bmatrix}=  	
 	\overbrace{
 	 	\begin{bmatrix}
 		I_1-I_2 & I_5-I_6 & -I_8 & -I_{10}\\
 		I_2-I_3 & I_6 & I_8 & I_{10}\\
 		I_3 & 0 & 0 & 0\\
 		I_4 & I_7 & I_9 & I_{11}
 	\end{bmatrix}}^{\mathtt{\hat M} }	\begin{bmatrix}
 	\mu \\
 	1 \\
 	1 \\
 	\mu 
 	\end{bmatrix}.
 \end{eqnarray}
 Define
 \begin{equation}
 	\mathtt{L}
 	=
 	\begin{bmatrix}
 		1 & 1 & -\dfrac{I_1-I_3}{I_3} & 0\\
 		0 & 0 & 1 & 0\\
 		0 & 1 & 0 & -\dfrac{I_{10}}{I_{11}}\\
 		0 & 0 & 0 & 1
 	\end{bmatrix}~~~\mbox{and}~~~\mathtt{P}
 	=
 	\begin{bmatrix}
 	0 & 1 & 0 & 0\\
 	1 & 0 & 0 & 0\\
 	0 & 0 & 1 & 0\\
 	0 & 0 & 0 & 1
 	\end{bmatrix}.
 \end{equation}
 Then
 \begin{equation}
 	\mathtt{L}\,\mathtt{\hat M}\,\mathtt{P}
 	=
 	\begin{bmatrix}
 		I_5 & 0 & 0 & 0\\
 		0 & I_3 & 0 & 0\\
 		I_6-\dfrac{I_{10}I_7}{I_{11}}
 		&
 		I_2-I_3-\dfrac{I_{10}I_4}{I_{11}}
 		&
 		I_8-\dfrac{I_{10}I_9}{I_{11}}
 		&
 		0\\
 		I_7 & I_4 & I_9 & I_{11}
 	\end{bmatrix}.
 \end{equation}
This means the first order variations of the weighted sum-rate correspond to the components of the vector
\begin{equation}
	\begin{bmatrix}
		I_5 & 0 & 0 & 0\\
		0 & I_3 & 0 & 0\\
		I_6-\dfrac{I_{10}I_7}{I_{11}}
		&
		I_2-I_3-\dfrac{I_{10}I_4}{I_{11}}
		&
		I_8-\dfrac{I_{10}I_9}{I_{11}}
		&
		0\\
		I_7 & I_4 & I_9 & I_{11}
	\end{bmatrix}	\begin{bmatrix} 1 \\ \mu \\ 1 \\ \mu	\end{bmatrix}
	\label{Eq179}
 \end{equation}
 Let us focus on the first two rows in \ref{Eq179}.
 Terms $I_5$ and  $I_3$ correspond  to compound random variables $\mathtt{n}_{\mathtt{u}_2}=\mathtt{s}_{\mathtt{u}_1}=\mathtt{u}_1+\mathtt{v}_1
 +\sqrt{b}\mathtt{v}_2+\mathtt{z}_1$
 and ${\mathtt{s}_{\mathtt{u}_2}}=\mathtt{u}_1+\mathtt{v}_1
 +\sqrt{b}\mathtt{u}_2+\sqrt{b}\mathtt{v}_2+\mathtt{z}_1$, respectively. Adding the constraints on power and area of each compound random variable to its corresponding term, and noting derivations in Appendices \ref{LIC2}, \ref{LIC1}, it is concluded that $\mathtt{u}_2$ and $\mathtt{u}_1+\mathtt{v}_1
 +\sqrt{b}\mathtt{v}_2+\mathtt{z}_1$ are Gaussian.  This means the signal, as well as the noise, in successive decoding of $\mathtt{u}_2$ at $Y_1$ are Gaussian\footnote{A primary difficulty in establishing the capacity region is to show that both signal and noise in additive noise channels corresponding to \ref{ApEq1ppnN} to \ref{ApEq4ppnN} should be Gaussian. Above arguments establish this result for \ref{ApEq1ppnN}.}.
 Noting  $\mathtt{u}_1+\mathtt{v}_1
 +\sqrt{b}\mathtt{v}_2+\mathtt{z}_1$ is Gaussian and $\mathtt{u}_1$, $\mathtt{v}_1$, $\mathtt{v}_2$, $\mathtt{z}_1$ are independent, by  applying Cramér’s theorem ~\cite{GN1}\cite{GN2}, it follows that $\mathtt{u}_1$, $\mathtt{v}_1$, $\mathtt{v}_2$ should be Gaussian.~~~~~$\square$
 
Next, Appendix \ref{LIC2} shows that the Gaussian distribution is the unique density for which the first-order variation of the entropy, subject to constraints on power and area under the density function, vanishes. 
%For the optimization problems defined by \ref{13-0} to \ref{NNEq7}, the optimized objective function is a linear combination of the mutual information terms appearing on the right-hand sides of the active constraints. Hence, the objective can be written as a linear combination of entropy terms associated with distinct compound random variables.
%Since LICQ is satisfied, the first-order variation of this linear combination must vanish at any maximizing solution. Because the compound random variables are distinct, this can occur only if the first-order variation associated with each entropy term vanishes individually. By Appendix \ref{LIC2}, this is possible only when each  compound random variable is Gaussian. 

\subsection{Generic Example for Functional Terms in \ref{ApEq1ppnN} to \ref{ApEq4ppnN}} \label{NA}
Functional $J(f_V,f_W)$ below  is a generic model for  additive-noise channels  encountered in successive decoding:
\begin{equation} 
	J(f_V,f_W)  =   H_1(V+W+Z)- H_2(W+Z) 
	\label{Eq228}
\end{equation}
where $V$ and $W$ are independent random variables with given power values, $H_1$, $H_2$  are entropy terms, and  $Z\sim \mathcal{N}(0,1)$. For simplicity, in verifying LICQ, $V$ and $W$ are selected as ``scalar" random variables with fixed power values $P_V$ and $P_W$. Generalization to vector code-books will be straightforward.   
%Note that the  pair of compound random variables corresponding to  the two entropy terms in  \ref{ApEq1ppnN}, \ref{ApEq2ppnN},\ref{ApEq3ppnN},\ref{ApEq4ppnN} each are  the sum of two independent random variables. Consequently,  as will become clear from derivations,  the functional in \ref{Eq228} will be  sufficient for verifying LICQ in conjunction  with \ref{NNEq6ttV} to \ref{NNEq7}, or equivalently \ref{ApEq1ppn}  to \ref{ApEq4ppn}. 
Let us define
\begin{eqnarray}
	S_1 &=&  V+W+Z \\
	S_2&=&  W+Z.
\end{eqnarray}		
We have 
\begin{eqnarray} \label{N148}
	f_{S_1}(s_1)\equiv (f_V * f_W * f_Z)(s_1)
	&= &
	\int_{-\infty}^{\infty}
	\int_{-\infty}^{\infty}
	f_V(v)f_W(w)f_Z(s_1-v-w)dw dv \\ \label{N149}
	&= &
	\int_{-\infty}^{\infty}
	f_V(v)
	\int_{-\infty}^{\infty}
	f_W(w)f_Z(s_1-v-w)dw dv \\ \label{N149p}
	f_{S_2}(s_2)\equiv (f_W * f_Z)(s_2)
	&= &
	\int_{-\infty}^{\infty}
	f_W(w)f_Z(s_2-w) dw. 
\end{eqnarray}
The power constraints are
\begin{eqnarray} \label{N150}
	E(S_1^2)\,\equiv\, P_{s_1}=1+P_V+P_W & = & 	1+\int v^2 f_V(v)\,dv+\int w^2 f_W(w)\,dw=\int s_1^2 f_{S_1}(s_1)ds_1\\  \label{N151}
	E(S_2^2)\,\equiv\, P_{s_2}=1+P_W  & = & 	1+\int w^2f_W(w)\,dw=\int s_2^2 f_{S_2}(s_2)ds_2.
\end{eqnarray}
The normalization constraints are
\begin{eqnarray}  \label{N152p} 
	\int f_{S_1}(s_1)ds_1 & = & 1 \\  \label{N153p}
	\int f_{S_2}(s_2)ds_2 &  = & 1~~~\mbox{or equivalently}  \\ \label{N152} 
	\int f_{V}(v)dv & = & 1 \\  \label{N153}
	\int f_{W}(w)dw&  = & 1.
\end{eqnarray}
Density functions are perturbed as
\begin{eqnarray} \label{N241}
	f_V(v)+\epsilon_V \Lambda_V(v) \\ \label{N243}
	f_W(w)+\epsilon_W \Lambda_W(w).
\end{eqnarray}
As an example, first order variation of  \ref{N149}  with respect to a perturbation of  $f_V$  is equal to 
\begin{eqnarray} \label{NN238}
	\epsilon_V\!	\int_v  \int_w \Lambda_V(v)f_W(w)f_Z(s_1-v-w)dwdv.
	\label{NN239}
\end{eqnarray}
Consider first order variations of $H_1$ and $H_2$ in \ref{Eq228} at points $s_1$ and $s_2$, respectively.  
Based on derivations similar to Appendix \ref{AppB}, first-order variation of $H_1(V+W+Z)$ with respect to a perturbation of  $f_V$, denoted as $\delta H_1[\Delta V]$ hereafter,  is equal to: 
\begin{eqnarray} \nonumber
	\delta H_1[\Delta V] = \\ \nonumber
	\epsilon_V\! \int_{s_1} \Big[\!\int_v  \int_w \! \Big(\Lambda_V(v)f_W(w)f_Z(s_1-v-w)dwdv\Big) \Big(1+\log \int_v\int_w f_V(v)f_W(w)f_Z(s_1-v-w)dw dv\Big)\Big]d s_1  =  \\ 
	\epsilon_V\!\int_{s_1}  \Big(1+
	\log f_{S_1}(s_1)\Big)\int_v  \int_w \Lambda_V(v)f_W(w)f_Z(s_1-v-w)dw\,dv\,d s_1. ~~~~~~
\end{eqnarray}
Likewise, first order variations of $H_1$ and $H_2$ with respect to $f_W$ are
\begin{eqnarray} 
	\delta H_1[\Delta W]& = & 	\epsilon_W\!\int_{s_1}  
	\Big(1+	\log f_{S_1}(s_1)\Big)\int_v  \int_w f_V(v)\Lambda_W(w)f_Z(s_1-v-w)dw\,dv\,d s_1 \\
	\delta H_2[\Delta W] & = & 	\epsilon_W\!\int_{s_2}  \Big(1+
	\log f_{S_2}(s_2)\Big) \int_w \Lambda_W(w)f_Z(s_2-w)dw\,d s_2.
\end{eqnarray}
Upon perturbation, normalization constraints and power constraints remain valid if:
\begin{eqnarray} \label{Eq181n}
	\int_{s_1} \int_v  \int_w \! \Lambda_V(v)f_W(w)f_Z(s_1-v-w)dw dv ds_1=0 \\
	\int_{s_1} \int_v  \int_w \! \Lambda_W(w) f_V(v) f_Z(s_1-v-w)dw ds_1=0 \\
	\int_{s_2} \int_w \! \Lambda_W(w) f_Z(s_2-w)dw ds_2=0 \\
	\int_{s_1} \int_v  \int_w \! s_1^2 \Lambda_V(v)f_W(w)f_Z(s_1-v-w)dw dv ds_1=0 \\
	\int_{s_1} \int_v  \int_w \! s_1^2 \Lambda_W(w) f_V(v) f_Z(s_1-v-w)dw ds_1=0 \\  \label{Eq186n}
	\int_{s_2} \int_w \! s_2^2 \Lambda_W(w) f_Z(s_2-w)dw ds_2=0.
\end{eqnarray}
LICQ requires the gradients of the active constraints\footnote{Even though LICQ focuses on the constraints, the form of the objective function determines which constraints among \ref{NNEq6ttV} to \ref{NNEq7} will be active. The optimum value of the weighted sum-rate will be a linear combination of the mutual information terms forming the right-hand sides of the active constraints, with positive coefficients.} to be linearly independent. Let us introduce linear coefficients $\alpha_V$, $\alpha_W$ as weights in forming these linear combinations. 
Noting \ref{Eq181n} to \ref{Eq186n},  the linear combinations in checking LICQ involves the perturbation of rate values only. 
LICQ fails if there exist perturbation functions 	$\Lambda_V(w), \Lambda_W(w)$   satisfying \ref{Eq181n} to \ref{Eq186n}  such that 
\begin{equation} \label{Eq188}
	\alpha_V\epsilon_V \delta H_1[\Lambda_V] +
	\alpha_W\epsilon_W \Big(\delta H_1[\Lambda_W]- H_2[\Lambda_W]\Big) = 0. 
\end{equation}
The above condition is satisfied if
\vspace{-0.4cm} 
\begin{eqnarray} \label{Eq189}
	\alpha_V \delta H_1[\Lambda_V]=0 & \Longrightarrow &  \overbrace{\alpha_V=0}^{(1)}~~\mbox{or}~~
	\overbrace{\delta H_1[\Lambda_V]=0}^{(2)}~\Longrightarrow~\Lambda_V(v)=0~\forall v~~~~~ \\ 	\label{Eq190}
	\alpha_W \Big(\delta H_1[\Lambda_W]- H_2[\Lambda_W]\Big) = 0 & \Longrightarrow & \overbrace{\alpha_W=0}^{(3)}~~\mbox{or}~~\overbrace{\delta H_1[\Lambda_W] = \delta H_2[\Lambda_W]}^{(4)}.
\end{eqnarray}
In \ref{Eq189}, condition (1) is aligned with LICQ being satisfied, while condition (2) is a trivial case and can be ignored. 
In \ref{Eq190}, condition (3) is aligned with LICQ being satisfied, and assuming $\Lambda_W(w)\neq 0$, condition (4) is equivalent to:
\begin{eqnarray}
	\int_{s_1} \Big(1+	\log f_{S_1}(s_1)\Big)\int_v  \int_w \Lambda_W(w) f_V(v)f_Z(s_1-v-w)dw\,dv\,ds_1  = \\
	\int_{s_2}  \Big(1+\log f_{S_2}(s_2)\Big) \int_w \Lambda_W(w)f_Z(s_2-w)dw\,d s_2.
\end{eqnarray}

\begin{eqnarray}
	I_1
	=
	\int_w
	\Lambda_W(w)
	A_1(w)
	\,dw
	& \mbox{where} &
	A_1(w)
	=
	\int_s
	\Big(1+\log f_{S_1}(s)\Big) f_{V+Z}(s-w)ds\\
	I_2
	=
	\int_w
	\Lambda_W(w)
	A_2(w)
	\,dw
	& \mbox{where} &
	A_2(w)
	=
	\int_s
	\Big(1+\log f_{S_2}(s)\Big)
	f_Z(s-w)
	\,ds.
\end{eqnarray}
This requires 
\begin{equation}
	I_1-I_2=0~\Longrightarrow~\int_w
	\Lambda_W(w)
	\Big(
	A_1(w)-A_2(w)
	\Big)
	\,dw
	=
	0.
\end{equation}
Noting \ref{Eq181n} to \ref{Eq186n} and assuming zero mean densities, the corresponding condition is
\begin{equation}
	A_1(w)-A_2(w)
	=
	c_0+c_2w^2.
	\label{Eq195}
\end{equation}
This is the Euler-Lagrange equation corresponding to the optimization of \ref{Eq228}. This  can happen in the following two cases
\begin{equation}
	\mbox{\underline{Case (1): first order variation of~\ref{Eq228}~\mbox{is zero}}}~~ \mbox{or}~~ \underline{\mbox{Case (2):}~\alpha_w=0}.
	\label{Eq196nn}
\end{equation}
Conditions in \ref{Eq196nn} entails for any feasible solution with non-zero first order variation, i.e.,  if Case (1) is violated, we should have  $\alpha_w$, i.e.,  the LICQ condition is satisfied, which in turn results in vanishing first order variation.  Overall, this means  \underline{any optimum  solution (local or global) should have vanishing first order variations}.

			\section{Proof of Theorem \ref{V4Th6} } \label{PrX13}
			
				{\bf Claim:} 
	{\em 
				Cardinality of the set $\Bar{\Bar{\mathsf{S}}}_{\upsilon}$ can be reduced, by increasing $\upsilon<\mu^s$, in a recursive manner, such that the final set is associated with a single $(\delta P_1,\delta P_2)$}. 

		\vspace{0.25cm}
		\noindent 
			{\bf Proof:}
		Let us assume the original set $\Bar{\Bar{\mathsf{S}}}_{\upsilon}$  is associated with $M$ distinct vectors $(\delta^m P_1,\delta^m P_2)$, $m=1,\ldots,M$. Each of these $M$ vectors is associated with a respective set of code-books' densities.  Consider  
		\begin{equation}
			(\breve{\delta} P_1,\breve{\delta} P_2)= (\min_{m}\delta^m P_1, \min_{m} \delta^m P_2).
			\label{Ae14}
		\end{equation} 
		The pair $(\breve{\delta} P_1,\breve{\delta} P_2)$ is called the Pareto minimal point corresponding to the set $(\delta^m P_1,\delta^m P_2)$, $m=1,\ldots,M$.
		Let us use  $(\breve{\delta} P_1,\breve{\delta} P_2)$ to compute new values for
		$(\mathbf{\Upsilon},\mathbf{\Gamma})$ and select the subset with smallest value of $\upsilon$ denoted as $\breve{\upsilon}$. Accordingly, let us form the sets $\Bar{\mathsf{S}}_{\breve{\upsilon}}$ and $\Bar{\Bar{\mathsf{S}}}_{\breve{\upsilon}}$. 
		Starting from the power reallocation vector $(\breve{\delta}P_1,\breve{\delta}P_2)$, each of the pairs $(\delta^m P_1,\delta^m P_2)$, $m=1,\ldots,M$, can be reached relying on a step with power reallocation $(\delta^m P_1-\breve{\delta}P_1,\delta^m P_2-\breve{\delta} P_2)$. This is possible since $\delta^m P_1-\breve{\delta}P_1\geq 0$ and $\delta^m P_2-\breve{\delta}P_2\geq 0$.  This means relying on  $\Bar{\Bar{\mathsf{S}}}_{\breve{\upsilon}}$ to achieve the next point on the boundary does not contradict the possibility of further moving counterclockwise to achieve 
		the boundary point corresponding to $\Bar{\Bar{\mathsf{S}}}_{\upsilon},~\upsilon<\breve{\upsilon}$. 
		Now let us shrink the range for power reallocation  vector by setting
		\begin{equation}
			{\Delta}\!P_1=\breve{\delta} P_1~~\mbox{and}~~{\Delta}\!P_2=\breve{\delta} P_2.
			\label{Eq54V4}	
		\end{equation}
		Accordingly, let us construct new sets following~\ref{Ae7} and \ref{Eq58New65}. Having multiple elements in  $\Bar{\Bar{\mathsf{S}}}_{\breve{\upsilon}}$ allows recursively moving in clockwise direction, where $\mathbf{\Upsilon}$ increases and $\mathbf{\Gamma}$ decreases in each step.  This procure can continue  until one of the following  cases occurs. Case (i): The value of  $\mathbf{\Gamma}$ at the final point is zero. 	Case~(ii):  The final set includes a single Pareto minimal power reallocation vector achieving a single point on the boundary. 	
		Case (i) entails no further counterclockwise step along the boundary is feasible, requiring a change in the strategy. 
		In Case (ii), from Theorem  \ref{V4Th6}, it follows that there is a Pareto minimal power reallocation which, in conjunction with zero-mean Gaussian code-books for compound random variables,   results in a unique point on the boundary. ~~~~$\square$

\end{appendix}

\end{document}